\documentclass[amsmath,amssymb,superscriptaddress,aps,reprint,widetext,longbibliography]{revtex4-2}

\usepackage[utf8]{inputenc}
\usepackage{graphicx}
\usepackage{xcolor}
\usepackage{hyperref}
\usepackage{notes2bib}
\usepackage{siunitx}
\usepackage{scalerel}
\usepackage{bm}
\usepackage{mathtools}
\usepackage{comment}
\newcommand{\smallx}{\scaleobj{.65}{{\mathcal{X}}}}

\begin{document}

\title{Identities for nonlinear memory kernels}

\author{Juliana Caspers}
\email{j.caspers@theorie.physik.uni-goettingen.de}
\affiliation{Institute for Theoretical Physics, Georg-August-Universit\"{a}t G\"{o}ttingen, 37073 G\"{o}ttingen, Germany}

\author{Matthias Krüger}
\email{matthias.kruger@uni-goettingen.de}
\affiliation{Institute for Theoretical Physics, Georg-August-Universit\"{a}t G\"{o}ttingen, 37073 G\"{o}ttingen, Germany}

\begin{abstract}
    Perturbing a system far away from equilibrium via a time dependent protocol can formally be described by a nonlinear Volterra series expansion. Here we derive identities for the nonlinear memory kernels arising in such nonlinear expansion, including the possibility of a nonlinear coupling between perturbation and system. 
    These identities rely on local detailed balance, and they include the fluctuation dissipation theorem as the lowest order identity. We test them in simulations for driven over- and underdamped Brownian particles. These identities for memory kernels can be recast in a series relation for the non-equilibrium cumulants of the observable conjugate to the driving and the observable described by the Volterra series. 
\end{abstract}

\maketitle

\section{Introduction}

When a system is driven away from equilibrium by a perturbation protocol, its non-equilibrium evolution is generally expected to be a convolution of the protocol with linear or nonlinear memory kernels~\cite{onsager_reciprocal_1931,kawasaki_theory_1968,evans_statistical_2008,fuchs_integration_2005,dhont_introduction_1996}. 
Such memory kernels have been found via projection operator techniques~\cite{zwanzig_memory_1961,mori_transport_1965,zwanzig_nonequilibrium_2001,grabert_projection_1982}, in which a  set of degrees of freedom are projected out, 
or in mode-coupling theory (MCT)~\cite{gotze_complex_2008, janssen_mode-coupling_2018}. Notably, in these formulations for  systems close to equilibrium, the memory kernels formally appear in response to thermal fluctuations rather than external perturbations~\cite{onsager_reciprocal_1931}. Such formulations then give rise to non-Markovian stochastic descriptions, including  non-Markovian Langevin equations, exemplified for the sub-diffusive plateau in mean squared displacements~\cite{fuchs_asymptotic_1998} observed experimentally~\cite{van_zanten_brownian_2000,lu_probe_2002,van_der_gucht_brownian_2003,caspers_how_2023}.
For linear systems, such as for a probe particle in a bath of harmonic oscillators, the famous Caldeira-Leggett   model~\cite{caldeira_influence_1981,zwanzig_nonequilibrium_2001}, explicit forms for the appearing kernels are found.

Driving non-linear systems more strongly will bring them far away from equilibrium causing nonlinear response and nonlinear fluctuations with less well known properties. Examples of behaviors that have been found for strongly driven systems are particle oscillations~\cite{doerries_correlation_2021,venturelli_memory-induced_2023}, shear-thinning for macro- or micro-shear~\cite{squires_simple_2005,gazuz_active_2009,harrer_force-induced_2012} or nonlinear behavior of Brownian particles immersed in micellar fluids~\cite{jayaraman_oscillations_2003,handzy_oscillatory_2004,berner_oscillating_2018,jain_two_2021}. Also superdiffusive behavior~\cite{harrer_force-induced_2012,winter_active_2012,benichou_geometry-induced_2013} can occur under strong driving. Nontrivial fluctuations of system far away from equilibrium have been rationalized in terms of temperatures that are different from ambient temperature, so called effective temperatures~\cite{wilson_small-world_2011,demery_driven_2019,cugliandolo_effective_2011,puglisi_temperature_2017}.

It is challenging to describe systems driven far away from equilibrium theoretically. One successful route extends the above mentioned projection operator techniques to stronger driving, ~\cite{fuchs_integration_2005,brader_first-principles_2008, gazuz_active_2009,gazuz_nonlinear_2013,gruber_active_2016,meyer_non-stationary_2017,te_vrugt_mori-zwanzig_2019,meyer_dynamics_2019,glatzel_interplay_2021,vroylandt_position-dependent_2022,schilling_coarse-grained_2022,netz_derivation_2024}, including the integration through transient technique~\cite{fuchs_integration_2005}. Also approaches in 
field theory~\cite{demery_driven_2019,venturelli_memory-induced_2023} have been extended to systems far away from equilibrium. Dynamical density functional theory~\cite{penna_dynamic_2003, rauscher_dynamic_2007,de_las_heras_velocity_2018,schmidt_power_2022,schilling_coarse-grained_2022}, including power functional theory, provides powerful tools. Lattice models~\cite{benichou_geometry-induced_2013,leitmann_nonlinear_2013,leitmann_time-dependent_2018} allow for analytical solutions in certain cases. Nonlinear response theory~\cite{asheichyk_brownian_2021,kruger_modified_2016,muller_brownian_2020} within path integral formalism~\cite{seifert_stochastic_2012,colangeli_meaningful_2011,basu_frenetic_2015,maes_response_2020} forms another promising route. Using such nonlinear response theory, Volterra series expansions for the mean response as well as for fluctuations have been formulated recently~\cite{caspers_nonlinear_2024}.   

In this manuscript, we present a general Volterra series expansion for a system driven by a time dependent protocol, based on the methods of  
Refs.~\cite{kruger_modified_2016,muller_brownian_2020,caspers_nonlinear_2024}.  
The Volterra series contains nonlinear memory kernels for the cumulants involving the observable conjugate to the driving protocol, and the observable the series is designed for. In contrast to previous work \cite{caspers_nonlinear_2024}, we lift the restriction to overdamped Brownian particles, so that in this manuscript the perturbation in general enters nonlinearly in entropy production. The main new result in this manuscript are however identities for the nonlinear memory kernels occurring in this series. These are found to rely on local detailed balance, and the lowest order identity is the fluctuation dissipation theorem. We show how they yield a series identity for non-equilibrium cumulants. The latter becomes the familiar expansion of the Boltzmann weight for a step like protocol. These identities are tested in  simulations of driven over or underdamped Brownian particles with  nonlinear interactions~\cite{muller_properties_2020,jain_two_2021}.

\section{System and Summary of identities}

Consider a system of $N$ stochastic degrees of freedom $y^{(i)}$ ($i=1,\dots,N$), abbreviated ${\bf y}$, 
 in contact with a thermal bath at temperature $T$, coupled to a deterministic control parameter $x_s$ with time index $s$, via the energy function
 $U(\mathbf{y}_s,x_s)$. The system is prepared in equilibrium at time $s=t_0$ with $x$ taking the value $x_{t_0}$, i.e., for $s\leq t_0$, the distribution of $\mathbf{y}$ is the Boltzmann distribution for  $U(\mathbf{y},x_{t_0})$.    
 $x_s$ follows the protocol $\smallx = (x_s,t_0 \leq s \leq t)$, driving the system away from equilibrium  for $s>t_0$. 
The derivative of $U$ with respect to $x_s$ defines the observable conjugate to the protocol, $F_s\equiv F(\mathbf{y}_s,x_s)\equiv \partial_{x_s}U(\mathbf{y}_s,x_s)$.
In case $x_s$ is a position, $F_s$ is (minus) the force acting on the control parameter. $x_s$ may however, depending on the form of $U$, represent other types of perturbation, e.g., external fields or forces. 

In the state out of equilibrium, the cumulants of $F_s$ and its correlations with another state observable $B_t = B(\mathbf{y}_t,x_t)$ acquire non-equilibrium forms. We allow $B$ to explicitly depend on $x_t$ to include the important case of $B_t\equiv F_t$.     We assume that the cumulants $\langle B_t;F_{t_1};\dots;F_{t_m}\rangle$ can be expanded in a Volterra series in $\dot x_s$~\cite{caspers_nonlinear_2024}, in which the following expansion  (memory) kernels appear\footnotetext[2]{It will be detailed below why the expansion is performed around the equilibrium state corresponding to  $x_t$.}~\cite{Note2} 
\begin{align}
\begin{split}
    \Gamma^{(m,n)}&_{t-s_1,\dots,t-s_n;t-t_1,\dots,t-t_m} \\&\coloneqq \frac{\beta^{m+1}}{n!} \frac{\delta^n}{\delta \dot{x}_{s_1}\cdots \delta \dot{x}_{s_n}} \langle B_t;F_{t_1};\dots;F_{t_m}\rangle \Big|_{\{x_s\}=x_t}.
    \end{split}
    \label{eq:Gammamn}
\end{align}
Here, $m+1$ denotes the order of cumulant, and $n$ marks the order of $\dot x$.
We will in this manuscript derive identities for these kernels. The first identity reads
\begin{align}
\Gamma_{t}^{(0,1)} &= \Gamma_{t}^{(1,0)} \label{eq:id1}.
\end{align}
Eq.~\eqref{eq:id1} connects the linear response of the mean ($m=0$, $n=1$) with the zeroth order (i.e. equilibrium) of the covariance ($m=1$, $n=0$). It is the familiar fluctuation dissipation theorem~\cite{callen_irreversibility_1951,kubo_fluctuation-dissipation_1966}. The other identities are, to our knowledge, not known in literature. The second reads, 
\begin{align}
    \Gamma_{s_1,s_2}^{(0,2)} &= \frac{1}{2} \sum_{\pi \in S_2} \Gamma_{s_{\pi(1)};s_{\pi(2)}}^{(1,1)}  - \frac{1}{2} \Gamma_{s_1,s_2}^{(2,0)} .\label{eq:id2}
    \end{align}
Here, the sum runs over all elements of the permutation group $S_2$.
Eq.~\eqref{eq:id2} is nonlinear, as it connects the second order response of the mean  ($m=0$, $n=2$) with the first order response of the covariance  ($m=1$, $n=1$) and the zeroth order of the third cumulant ($m=2$, $n=0$). The appearance of the third cumulant is remarkable, as, this way, Eq.~\eqref{eq:id2} connects nonlinear response to non-Gaussian fluctuations. The next identity reads 
        \begin{align}
        \begin{split}
       \Gamma_{s_1,s_2,s_3}^{(0,3)} &= \frac{1}{6} \sum_{\pi \in S_3} \Gamma_{s_{\pi(1)},s_{\pi(2)};s_{\pi(3)}}^{(1,2)} \\
       &\quad- \frac{1}{12} \sum_{\pi \in S_3} \Gamma_{s_{\pi(1)};s_{\pi(2)},s_{\pi(3)}}^{(2,1)} + \frac{1}{6} \Gamma_{s_1,s_2,s_3}^{(3,0)} 
       \end{split}
    \label{eq:id3} 
   \end{align}
with the sum over all elements in the permutation group $S_3$.
Eq.~\eqref{eq:id3} connects the third order response of mean  ($m=0$, $n=3$), with the second order response of the covariance ($m=1$, $n=2$), the first order response of the third cumulant ($m=2$, $n=1$) and the zeroth order of the fourth cumulant ($m=3$, $n=0$). While Eqs.~\eqref{eq:id1} - \eqref{eq:id3} suggest a continuation to higher orders, we have only explicitly verified the three identities given above. We speculate about the form of higher orders in the conclusion section.

These identities rely on local detailed balance~\cite{maes_local_2021} as will be detailed below. It is worth noticing that 
$F$ and $B$ may range from micro- to macroscopic scales, as they can, e.g., be the positions of molecular particles, or macroscopic observables such as the macroscopic magnetization.  The mentioned identities thus appear to be applicable on various length scales.

\section{Derivation of Identities}\label{chap:LDB}

We will lay out the derivation of the above identities in this section. 
Notably, as the potential $U$ contains the protocol $x_s$ in arbitrary nonlinear order, this includes an explicit expansion of the conjugate observable $F$ in $\dot x_s$, and the corresponding expansion of the  entropic part of the action.

\subsection{Thermodynamic considerations}\label{chap:NonlinearFDT_Generalization_Setup}

Due to the time dependence of $x_s$, the energy function $U$ depends explicitly on time for $s\geq t_0$. It is advantageous to split $U$ into the equilibrium part acquired at time $s\leq t_0$ and its deviation, 
\begin{align}
    U(\mathbf{y}_s,x_s) =  U(\mathbf{y}_s,x_{t_0})+
     \underbrace{ U(\mathbf{y}_s,x_s) - U(\mathbf{y}_s,x_{t_0})}_{H_1(\mathbf{y}_s,x_s,x_{t_0})}.
    \label{eq:TimeDependentHamiltonian}
\end{align}
The perturbed, protocol-dependent part of the energy function is denoted by $H_1$, it plays the role of the perturbation Hamiltonian; It yields the change of  $U(\mathbf{y}_s,x_s)$ with respect to the reference equilibrium energy, where the protocol is fixed at $x_{t_0}$.

To determine the action of the perturbed process, we  consider the  change  $\Delta U$ between times $t_0$ and $t$. From the first law of thermodynamics,
\begin{align}
   \Delta U= U(\mathbf{y}_s,x_s)\Big|_{s=t_0}^t = W-Q.
    \label{eq:1stlaw_thermodyn}
\end{align}
$Q$ denotes the dissipated heat, i.e., energy distributed in the bath in contact with the degrees ${\bf y}$, and  $W$ refers to the work applied,
\begin{align}
    W = \int_{t_0}^t \mathrm{d}s\, \dot{x}_s \frac{\partial U}{\partial x_s} = \int_{t_0}^t \mathrm{d}s\, \dot{x}_s F(\mathbf{y}_s,x_s).
    \label{eq:WorkonSystem}
\end{align}
From the dissipated heat we compute the entropy change $\mathcal{S} = \beta Q$ of the system of degrees ${\bf y}$,
\begin{align}
    \mathcal{S} &= \beta \left( -\Delta U + W \right)\notag \\
    &=\beta \left[ U(\mathbf{y}_{t_0},x_{t_0})-U(\mathbf{y}_t,x_t) + \int_{t_0}^t \mathrm{d}s\, \dot{x}_s F(\mathbf{y}_s,x_s)  \right].
    \label{eq:SHamiltonianAnsatz2}
\end{align}
Entropy production in Eq.~\eqref{eq:SHamiltonianAnsatz2} contains a part from the unperturbed potential $U({\bf y}_s,x_{t_0})$  and a part from $H_1(\mathbf{y}_s,x_s,x_{t_0})$ in Eq.~\eqref{eq:TimeDependentHamiltonian}. Only the latter enters the Volterra series below, denoted $S$,
\begin{align}
     S&\equiv \mathcal{S}-  \mathcal{S}|_{\{x_s\}=x_{t_0}}\notag\\
      &= \beta \left( -\Delta H_1+ W \right)\notag \\
    &=\beta \left[ U(\mathbf{y}_{t},x_{t_0})-U(\mathbf{y}_t,x_{t}) + \int_{t_0}^t \mathrm{d}s\, \dot{x}_s F(\mathbf{y}_s,x_s)  \right].
    \label{eq:SHamiltonianAnsatz}
\end{align}

\subsection{Volterra series expansion}\label{chap:NonlinearFDT_Generalization_Volterra}

We aim to expand the cumulants of observables $B(\mathbf{y}_t,x_t)$ and $F(\mathbf{y}_{t_i},x_{t_i})$ in a Volterra series in powers of the control parameter, around the equilibrium state with the control parameter being fixed at $x_{t_0}$~\cite{caspers_nonlinear_2024}. We will see that this expansion takes a power series in $\dot x_s$, i.e., the time derivative of $x_s$. 
We consider a path integral representation, where $\omega = (\mathbf{y}_s,t_0 \leq s \leq t)$ denotes the path of the stochastic degrees of freedom $\mathbf{y}$ over the time interval $[t_0,t]$. Recall the protocol (path)  $\smallx = (x_s,t_0 \leq s \leq t)$.
The expectation value of an observable $O(\smallx,\omega)$  is given by
\begin{align}
    \langle O \rangle = \int \mathcal{D}\omega \, e^{-\mathcal{A}(\smallx,\omega)}\mathcal{P}_\mathrm{eq}(x_{t_0},\omega)O(\smallx,\omega).\label{eq:PathintegralProtocolDependent}
\end{align} 
$\mathcal{P}_\mathrm{eq}(x_{t_0},\omega)$ denotes the path weight of the corresponding equilibrium state, and $\mathcal{A}$ is the action quantifying the change due to the perturbation. $O$ will be the corresponding products of $B_t$ and $F_{t_i}$, to obtain the sought cumulants in Eq.~\eqref{eq:Gammamn}.

We decompose $\mathcal{A}$ into a time-symmetric and a time-antisymmetric component~\cite{baiesi_fluctuations_2009,maes_response_2020}. 
The basic assumption employed concerns the connection of the antisymmetric part with the entropy production of Eq.~\eqref{eq:SHamiltonianAnsatz}, i.e.,
\begin{align}
    S &=\ln \frac{\mathcal{P}_\mathrm{eq}(x_{t_0},\omega)e^{-\mathcal{A}(\smallx,\omega)}}{\mathcal{P}_\mathrm{eq}(x_{t_0},\theta\omega)e^{-\mathcal{A}(\theta\smallx,\theta\omega)}}\label{eq:db}\\&= \mathcal{A}(\theta\smallx,\theta \omega)-\mathcal{A}(\smallx,\omega),\label{eq:sl}
\end{align}
with time reversal operator $\theta$, i.e.,  $(\theta \smallx,\theta \omega)_s = (\pi x_{t+t_0-s},\pi\mathbf{y}_{t+t_0-s})$, where $\pi$ denotes the kinematic time-reversal flipping the sign of velocities. The  assumption formulated in Eq.~\eqref{eq:db} is known as the condition of \textit{local detailed balance}\footnotetext[1]{Local detailed balance amounts to assuming that all hidden degrees of freedom are equilibrated and in contact with heat baths.
Consider, for example, a driven Brownian particle in a fluid.
In a good approximation, the fluid molecules remain in thermal equilibrium~\cite{falasco_local_2021}. 
For explanations and counter-examples of systems where local detailed balance is broken, the reader is referred to Ref.~\cite{maes_local_2021}.}~\cite{Note1,maes_local_2021}.
In Appendix~\ref{app:EntropicComponentGeneralLangevin} we provide an example of underdamped Langevin dynamics  for which the equality of the expressions in  Eq.~\eqref{eq:SHamiltonianAnsatz} and Eq.~\eqref{eq:db} is found from the Onsager-Machlup action. Eq.~\eqref{eq:sl} uses  $\mathcal{P}_\mathrm{eq}(x_{t_0},\theta\omega)=\mathcal{P}_\mathrm{eq}(x_{t_0},\omega)$, i.e., symmetry of equilibrium path weights. We denote $D = \frac{1}{2}\left(\mathcal{A}(\theta\smallx,\theta \omega)+\mathcal{A}(\smallx,\omega)\right)$, the symmetric part.

The expansion is  cumbersome due to (i) the boundary terms in Eq.~\eqref{eq:SHamiltonianAnsatz}, (ii) the nonlinear dependence of $S$ on $x_s$, and (iii) the dependence of $F$ on $x_s$. A simplification occurs by taking the limit of $t_0 \to -\infty$ and setting the initial value of the control parameter equal to the final one, $x_{t_0}=x_t$~\cite{kruger_modified_2016}. 
As $x_{t_0}$ is the value in the infinite past, this choice has no influence on the system at time $t$. One has however to keep in mind that the expansion is performed around the equilibrium state corresponding to $x_t$, and this defines the kernels in Eq.~\eqref{eq:Gammamn}. With this choice, the first two terms in Eq.~\eqref{eq:SHamiltonianAnsatz} cancel, and $S$ simplifies to
\begin{align}
    S = \beta W = \frac{1}{k_B T} \int_{-\infty}^t \mathrm{d}s\, \dot{x}_s F(\mathbf{y}_s,x_s). \label{eq:SHamiltoniant0Inf}
\end{align}
As mentioned, the stochastic conjugate observable $F$ explicitly depends on $x_s$, rendering $S$ nonlinear in $\dot{x}$. 
This is seen by expanding $F_s$ in powers of $(x_s-x_{t_0})$ recalling that $x_{t_0}=x_t$,
\begin{align}
\begin{split}
    F(\mathbf{y}_s,x_s) &= F(\mathbf{y}_s,x_t) + (x_s-x_t) F'(\mathbf{y}_s,x_t) \\
    &\quad+ \frac{1}{2} (x_s-x_t)^2 F''(\mathbf{y}_s,x_t) + \dots 
\end{split}\label{eq:forceexpansion}
\end{align}
The primes denote derivatives with respect to $x_s$, which are evaluated at $x_s=x_t$.
The resulting expansion of $S = S' + S''/2+ S'''/6+\dots$ contains the terms
\begin{subequations}\label{eq:Ss}
\begin{align}
    S' &= \frac{1}{k_B T} \int_{-\infty}^t \mathrm{d}s \, \dot{x}_s F(\mathbf{y}_s,x_t), \\
    S'' &= -\frac{2}{k_B T} \int_{-\infty}^t \mathrm{d}s \int_{-\infty}^t \mathrm{d}s'\, \dot{x}_s \dot{x}_{s'} F'(\mathbf{y}_s,x_t) \Theta(s'-s) ,\\
    \begin{split}
    S''' &= \frac{3}{k_B T} \int_{-\infty}^t \mathrm{d}s_1 \int_{-\infty}^t \mathrm{d}s_2 \int_{-\infty}^t \mathrm{d}s_3\, \dot{x}_{s_1} \dot{x}_{s_2} \dot{x}_{s_3}\\
    &\quad\quad\quad\quad\times F''(\mathbf{y}_{s_1},x_t) \Theta(s_2-s_1)\Theta(s_3-s_1) .
    \end{split}\\
&\;\;\vdots \nonumber
\end{align}
\end{subequations}
We used the identity $x_s-x_t = -\int_{-\infty}^t \mathrm{d}s'\, \dot{x}_{s'}\Theta(s'-s)$ with the Heaviside function $\Theta(t)$ to make the dependence on $\dot x$ explicit. In the often considered case that $U$ is linear in $x_s$~\cite{basu_frenetic_2015, holsten_thermodynamic_2021}, only $S'$ in the above is finite, with $S''$, $S'''$, $\dots$ vanishing.  

The time-symmetric part $D$ of the action in general takes no such simple form as it depends on dynamical details. This however poses no problem, as the form of $D$ is not needed to find the derived identities. 
To get there, we introduce notation $D= D'+D''/2+\dots$ with
\begin{align}
    D' &= \int_{-\infty}^t \mathrm{d}s\, \dot{x}_s \mathcal{D}_s \label{eq:D1general}\\
    D'' &= \int_{-\infty}^t \mathrm{d}s \int_{-\infty}^t \mathrm{d}s'\, \dot{x}_s \dot{x}_{s'} \mathcal{D}_{s,s'}\label{eq:D2general} ,\\
    &\;\;\vdots\notag
\end{align}
and with $\mathcal{D}_s$, $\mathcal{D}_{s,s'}$, \dots, as mentioned, generally unknown coefficients.

By expanding also the exponential in Eq.~\eqref{eq:PathintegralProtocolDependent} (compare calculation steps in Ref.~\cite{caspers_nonlinear_2024}), we find for the response of any path observable $O = O(\smallx,\omega)$, 
\begin{align}
\begin{split}
    \langle O \rangle  &= \langle O\rangle_\mathrm{eq} + \frac{1}{2} \langle S';O\rangle_\mathrm{eq} - \langle D';O\rangle_\mathrm{eq}\\
    &\quad+ \frac{1}{2} \left[ \langle D';D';O\rangle_\mathrm{eq}  - \langle D';S';O\rangle_\mathrm{eq} \right]\\
    &\quad- \frac{1}{2}  \langle D'';O\rangle_\mathrm{eq} + \frac{1}{4} \langle S'';O\rangle_\mathrm{eq} + \frac{1}{8} \langle S';S';O\rangle_\mathrm{eq}+ \dots.
\end{split}
\label{eq:ResponseOConnectedCorrGeneralized}
\end{align}
$\langle \cdot \rangle_\mathrm{eq}$ denote equilibrium averages with the protocol fixed at $x_t$,
\begin{align}
        \langle O \rangle_\mathrm{eq} = \int \mathcal{D}\omega \, \mathcal{P}_\mathrm{eq}(x_{t_0},\omega)O(\smallx,\omega).\label{eq:EqAverage}
\end{align}
However, the path observables in Eq.~\eqref{eq:ResponseOConnectedCorrGeneralized} depend explicitly on the non-equilibrium protocol $x_s$.
To arrive at an explicit Volterra series expansion of the response in $\dot x$ the terms in Eq.~\eqref{eq:ResponseOConnectedCorrGeneralized} are additionally expanded to obtain the desired orders in $\dot x$.

The response of a state observable $O_t = O(x_t,\mathbf{y}_t)$, in turn, allows for simplifications.
We define the expectation value of a state observable $O_t$ under the reversed path and protocol (compare Refs.~\cite{basu_frenetic_2015,caspers_nonlinear_2024}),
\begin{align}
    \langle (O\theta)_t\rangle &\equiv \int \mathcal{D}\omega e^{-\mathcal{A}(\theta\smallx,\theta\omega)} \mathcal{P}_\mathrm{eq}(x_{t_0},\theta\omega) O(x_t,\mathbf{y}_t)
    \label{eq:PathintegralOReversed}
\end{align}
and consider the difference of Eq.~\eqref{eq:ResponseOConnectedCorrGeneralized} and a respective expansion of Eq.~\eqref{eq:PathintegralOReversed}.
Because $\mathbf{y}_t$ correspond to the end of paths, which, in the reversed weights, corresponds to the distribution of equilibrium, one finds 
\begin{align}
    \langle (O\theta)_t\rangle 
    =\frac{1}{Z}\int \mathrm{d}\mathbf{y}\, e^{-\beta U(\mathbf{y},x_{t_0})}O(x_t,\mathbf{y})= \langle O_t\rangle_\mathrm{eq}
    \label{eq:PathintegralOReversed2}
\end{align}
with partition sum $Z$. In other words, $\langle (O\theta)_t\rangle$ evaluates to an equilibrium average, 
and we find
\begin{align}
\begin{split}
    \langle O_t \rangle   &= \langle O_t\rangle_\mathrm{eq} +  \langle S';O_t\rangle_\mathrm{eq} + \frac{1}{2} \langle S'';O_t\rangle_\mathrm{eq} - \langle D';S';O_t\rangle_\mathrm{eq} \\
    &\quad-\frac{1}{2}\left[ \langle D';S'';O_t\rangle_\mathrm{eq} + \langle D'';S';O_t\rangle_\mathrm{eq}\right]\\
    &\quad + \frac{1}{2} \langle D';D';S';O_t \rangle_\mathrm{eq} + \frac{1}{24} \langle S';S';S';O_t\rangle_\mathrm{eq} \\
    &\quad+ \frac{1}{3} \langle S''';O_t\rangle_\mathrm{eq}+ \dots
\end{split}
\label{eq:ResponseStateOConnectedCorrGeneralized}
\end{align}
These simplify further as $x_t=x_{t_0}$ is used. 
The presented form of Eqs.~\eqref{eq:ResponseOConnectedCorrGeneralized} and \eqref{eq:ResponseStateOConnectedCorrGeneralized} in terms of connected correlation functions follows from the list of identities, (these are modified compared to the ones given in Ref.~\cite{caspers_nonlinear_2024} due to the higher order derivatives of $S$),
\begin{align}
    \langle S'\rangle_\mathrm{eq} = \langle D'\rangle_\mathrm{eq} &= 0,\\
    \langle S''/2-D'S'\rangle_\mathrm{eq} &= 0,\\
    \langle D'^2S'-D'S''-D''S'+S'^3/12+S'''/3\rangle_\mathrm{eq} &= 0,\\
    \langle D''-D'^2-S'^2/4\rangle_\mathrm{eq} &= 0.
\end{align}

We will apply Eqs.~\eqref{eq:ResponseOConnectedCorrGeneralized} and \eqref{eq:ResponseStateOConnectedCorrGeneralized} to calculate the cumulants of the observables $B(\mathbf{y}_t,x_t)$ and $F(\mathbf{y}_{t_i},x_{t_i})$.
The mean is calculated using Eq.~\eqref{eq:ResponseStateOConnectedCorrGeneralized}, while higher-order cumulants are calculated using Eq.~\eqref{eq:ResponseOConnectedCorrGeneralized}.
We 
 find the following form of the Volterra series,
\begin{subequations}\label{eq:Volterra}
\begin{align}
\begin{split}
   &\beta \langle B_t\rangle = \Gamma^{(0,0)}+ \int_{-\infty}^t \mathrm{d}s\, \dot{x}_{s} \Gamma_{t-s}^{(0,1)} \\
   &\;+  \int_{-\infty}^t  \mathrm{d}s\int_{-\infty}^t  \mathrm{d}s'\,  \dot{x}_{s}\dot{x}_{s'} \Gamma_{t-s,t-s'}^{(0,2)}\\
   &\;+  \int_{-\infty}^t   \mathrm{d}s\int_{-\infty}^t   \mathrm{d}s' \int_{-\infty}^t   \mathrm{d}s'' \,\dot{x}_{s} \dot{x}_{s'} \dot{x}_{s''} \Gamma_{t-s,t-s',t-s''}^{(0,3)} + \dots 
  \end{split}
   \label{eq:MeanForceG}\\ 
       \begin{split}
       &\beta^2 \langle B_t;F_{t'}\rangle =  \Gamma_{t-t'}^{(1,0)} + \int_{-\infty}^t   \mathrm{d}s \, \dot{x}_{s} \Gamma^{(1,1)}_{t-s;t-t'} \\
       &\quad + \int_{-\infty}^t  \mathrm{d}s \int_{-\infty}^t  \mathrm{d}s' \, \dot{x}_{s} \dot{x}_{s'} \Gamma^{(1,2)}_{t-s,t-s';t-t'} + \dots \label{eq:ForceCovG} 
       \end{split}\\
\begin{split}
       &\beta^3 \langle B_t;F_{t'};F_{t''}\rangle = \Gamma^{(2,0)}_{t-t',t-t''} \\
       &\quad+ \int_{-\infty}^t  \mathrm{d}s \, \dot{x}_{s} \Gamma^{(2,1)}_{t-s;t-t',t-t''} 
       +\dots
\end{split}       
\label{eq:ForceCum3G} \\
        &\beta^4 \langle B_t;F_{t'};F_{t''};F_{t'''}\rangle = \Gamma^{(3,0)}_{t-t',t-t'',t-t'''} + \dots \label{eq:ForceCum4G}\\
                &\;\;\vdots  \nonumber
    \end{align}
    \end{subequations}
The memory kernels in these expansions are in accordance with the definition in Eq.~\eqref{eq:Gammamn}. The explicit microscopic expressions in terms of $S$ and $D$ are  given in Appendix~\ref{sec:memory}. 
The kernels $\Gamma^{(m,n)}_{t-s_1,\dots,t-s_n;t-t_1,\dots,t-t_m}$ are by construction symmetric in time arguments $s_1,\dots,s_n$ and symmetric in time arguments $t_1,\dots,t_m$.
In the identities below, we furthermore symmetrize them by hand  with respect to interchanges of time arguments $s_n$ and $t_m$. This symmetrization is needed for the indentities to hold. It poses no problem in experimental use of the identities, and the symmetrization also occurs naturally when representing the identities in integral form (see Eqs.~\eqref{eq:cumulants}, \eqref{eq:NonlinearFDTmean} and \eqref{eq:NonlinearFDTConvolvedForceCov} below). We repeat the identities of Eqs.~\eqref{eq:id1}-\eqref{eq:id3} for convenience,  the derivation follows from the kernel expressions given in Appendix \ref{sec:memory},
\begin{align*}
\Gamma_{t}^{(0,1)} &= \Gamma_{t}^{(1,0)} \tag{\ref{eq:id1}}\\
    \Gamma_{s_1,s_2}^{(0,2)} &= \frac{1}{2} \sum_{\pi \in S_2} \Gamma_{s_{\pi(1)};s_{\pi(2)}}^{(1,1)}  - \frac{1}{2} \Gamma_{s_1,s_2}^{(2,0)} \tag{\ref{eq:id2}}\\
    \begin{split}
    \Gamma_{s_1,s_2,s_3}^{(0,3)} &= \frac{1}{6} \sum_{\pi \in S_3} \Gamma_{s_{\pi(1)},s_{\pi(2)};s_{\pi(3)}}^{(1,2)}  + \frac{1}{6} \Gamma_{s_1,s_2,s_3}^{(3,0)} \\
    &\quad  - \frac{1}{12} \sum_{\pi \in S_3} \Gamma_{s_{\pi(1)};s_{\pi(2)},s_{\pi(3)}}^{(2,1)}
    \end{split}\tag{\ref{eq:id3}} \\
    &\;\;\,\vdots \nonumber
\end{align*}
We repeat that one may expect the series of identities to extend to higher orders, but, due to the cumbersome nature of derivation, we have only verified them up to the order given.  We note some similarity of these expressions to the relations found in Ref.~\cite{holsten_thermodynamic_2021}.

We also repeat that $B$ and $F$ can be macroscopic, such as, e.g., the force acting on the wall of a piston, the force acting on a colloidal particle. $F$ and $B$ can also be microscopic, e.g., one of the degrees ${\bf y}_t$.

Notably, each identity connects kernels with the same sum $m+n$.

\section{Interpretation of the identities} 
The identities of  Eqs.~\eqref{eq:id1} - \eqref{eq:id3} can be formulated in a relation between the cumulants of Eq.~\eqref{eq:Volterra},
\begin{align}
\begin{split}
    \beta\langle B_t \rangle&=\beta\langle B \rangle_\mathrm{eq} + \beta^2\int_{-\infty}^t \mathrm{d}s\,\dot x_s  \langle B_t;F_s \rangle\\
    &\quad- \frac{\beta^3}{2} \int_{-\infty}^t \mathrm{d}s \int_{-\infty}^t \mathrm{d}s'\,\dot x_s  \dot x_{s'}\langle B_t;F_s;F_{s'} \rangle\\
    &\quad+ \frac{\beta^4}{6} \int_{-\infty}^t \mathrm{d}s \int_{-\infty}^t\mathrm{d}s' \int_{-\infty}^t \mathrm{d}s''\, \dot x_s  \dot x_{s'}\dot x_{s''}\\
    &\quad \quad \times \langle B_t;F_s;F_{s'};F_{s''} \rangle
    +\mathcal{O}(\dot x^4).
    \end{split}
    \label{eq:cumulants}
\end{align}
Eq.~\eqref{eq:cumulants} contains the non-equilibrium cumulants. 
Eq.~\eqref{eq:cumulants}, as written, is valid up to order $\mathcal{O}(\dot x^4)$; this means that only the equilibrium part of the fourth cumulant enters, while the third cumulant enters up to linear order, and so on. We remind that Eq.~\eqref{eq:cumulants} is derived under the choice $x_{t_0}=x_t$, so that, e.g., $\beta\langle B \rangle_\mathrm{eq}$ is evaluated in the equilibrium ensemble with protocol value $x_{t}$. While we have formally expanded in powers of $\dot x$, the structure of Eq.~\eqref{eq:cumulants} suggests that the small dimensionless expansion parameter can be estimated to $\beta F \int_{t-\tau}^t \mathrm{d}s\, \dot x_s$ with relaxation time $\tau$ of the cumulants.   

Eq.~\eqref{eq:cumulants} shows that the identities allow to express any cumulant purely in terms of other cumulants. For example, the first cumulant can be expanded in terms of the following expansion coefficients \textcolor{black}{(memory kernels)} 
\begin{align}
\begin{split}
     \beta \langle B_t\rangle &= \langle B\rangle_\mathrm{eq} + \int_{-\infty}^t\mathrm{d}s\, \dot{x}_s \Gamma_{t-s}^{(1,0)} \\
     &\quad +  \int_{-\infty}^t \mathrm{d}s \int_{-\infty}^t \mathrm{d}s'\, \dot{x}_s \dot{x}_{s'} \left(\Gamma_{t-s,t-s'}^{(1,1)} -\frac{1}{2}\Gamma_{t-s,t-s'}^{(2,0)} \right)\\
    &\quad+ \int_{-\infty}^t \mathrm{d}s \int_{-\infty}^t \mathrm{d}s' \int_{-\infty}^t \mathrm{d}s''\, \dot{x}_s \dot{x}_{s'} \dot{x}_{s''}\Big( \Gamma_{t-s,t-s',t-s''}^{(1,2)} \\
    &\quad \quad+ \frac{1}{6} \Gamma_{t-s,t-s',t-s''}^{(3,0)} -\frac{1}{2} \Gamma_{t-s;t-s',t-s''}^{(2,1)} \Big)+ \mathcal{O}(\dot{x}^4) .
    \end{split}
\label{eq:NonlinearFDTmean}
\end{align}
In Eq.~\eqref{eq:NonlinearFDTmean}, the kernels of second, third, and fourth order cumulants appear on the right hand side. 
Analogously, the expansion of the second cumulant is given by
\begin{align}
\begin{split}
     &\beta^2 \int_{-\infty}^t \mathrm{d}s\, \dot{x}_s \, \langle B_t;F_s\rangle = \int_{-\infty}^t\mathrm{d}s\, \dot{x}_s \Gamma_{t-s}^{(0,1)} \\
     &\quad+  \int_{-\infty}^t \mathrm{d}s \int_{-\infty}^t \mathrm{d}s'\, \dot{x}_s \dot{x}_{s'} \left(\Gamma_{t-s,t-s'}^{(0,2)} +\frac{1}{2}\Gamma_{t-s,t-s'}^{(2,0)} \right)\\
    &\quad+ \int_{-\infty}^t \mathrm{d}s \int_{-\infty}^t \mathrm{d}s' \int_{-\infty}^t \mathrm{d}s''\, \dot{x}_s \dot{x}_{s'} \dot{x}_{s''}\Big( \Gamma_{t-s,t-s',t-s''}^{(0,3)} \\
    &\quad\quad - \frac{1}{6} \Gamma_{t-s,t-s',t-s''}^{(3,0)} +\frac{1}{2} \Gamma_{t-s;t-s',t-s''}^{(2,1)} \Big)+ \mathcal{O}(\dot{x}^4) .
\end{split}\label{eq:NonlinearFDTConvolvedForceCov}
\end{align}
Eq.~\eqref{eq:NonlinearFDTConvolvedForceCov} quantifies non-equilibrium fluctuations of $B$ and $F$. It is expressed in terms of mean of $B$, and third and fourth cumulants. 

Eqs.~\eqref{eq:NonlinearFDTmean} and \eqref{eq:NonlinearFDTConvolvedForceCov} also allow to determine the 
difference of the mean of $B$ and the covariance of $B$ and $F$,
\begin{align}
\begin{split}
    \beta \langle &B_t\rangle - \beta^2 \int_{-\infty}^t \mathrm{d}s\, \dot{x}_s \, \langle B_t;F_s\rangle \\
    &= \Gamma^{(0,0)}- \frac{1}{2} \int_{-\infty}^t \mathrm{d}s \int_{-\infty}^t \mathrm{d}s'\, \dot{x}_s \dot{x}_{s'} \Gamma_{t-s,t-s'}^{(2,0)}\\
    &\quad+ \frac{1}{2} \int_{-\infty}^t \mathrm{d}s \int_{-\infty}^t \mathrm{d}s' \int_{-\infty}^t \mathrm{d}s''\, \dot{x}_s \dot{x}_{s'} \dot{x}_{s''} \\
    &\quad\quad\quad \times\left( \frac{1}{3} \Gamma_{t-s,t-s',t-s''}^{(3,0)} - \Gamma_{t-s;t-s',t-s''}^{(2,1)} \right) + \mathcal{O}(\dot{x}^4) 
\end{split}\label{eq:DifferenceMeanForceconvolvedForceCov} \\
\begin{split}
    &= \beta\langle B \rangle_\mathrm{eq} - \frac{\beta^3}{2} \int_{-\infty}^t \mathrm{d}s \int_{-\infty}^t \mathrm{d}s'\,\dot x_s  \dot x_{s'}\langle B_t;F_s;F_{s'} \rangle\\
    &\quad+ \frac{\beta^4}{6} \int_{-\infty}^t \mathrm{d}s \int_{-\infty}^t\mathrm{d}s' \int_{-\infty}^t \mathrm{d}s''\, \dot x_s  \dot x_{s'}\dot x_{s''}\\
    &\quad \quad \times \langle B_t;F_s;F_{s'};F_{s''} \rangle
    +\mathcal{O}(\dot x^4).
\end{split}
\end{align}
The left hand side vanishes identically up to first order, as it resembles  FDT. The right hand side thus quantifies the deviation from FDT, starting  at second order. The first term on the right, notably, is the equilibrium third cumulant. 

Notably, the right hand side of Eq.~\eqref{eq:DifferenceMeanForceconvolvedForceCov} vanishes  for Gaussian observables $F$ and $B$, as then  the higher cumulants vanish. In such systems, the mean thus equals the integrated covariance to the given order. It will be interesting to investigate this for higher orders than the ones given here.

\section{Test: Driven probe particle in nonlinear bath}
In this section we will test the identities for $B \equiv F$ in simulations of a model of a driven probe particle in a nonlinear surrounding.

\subsection{Model}
We consider two coupled Brownian particles. Particle 1, the probe,  with position $y^{(1)}$ and a second particle, making up a nonlinear surrounding, with position $y^{(2)}$.
The particles carry friction coefficients $\gamma_1$ and $\gamma_2$, respectively, and masses $m_1$ and $m_2$. Varying the latter allows to test overdamped and underdamped cases. The particles follow the Langevin equations 
\begin{align}
    m_i \ddot{y}_t^{(i)}+\gamma_i \dot{y}_t^{(i)} &= -\partial_{y^{(i)}_t}U(\{y^{(i)}_t\},x_t) +\xi_t^{(i)} \label{eq:MasterLangevin}
\end{align}
with Gaussian white noises $\xi_t^{(i)}$,
\begin{align}
    \left\langle \xi_t^{(i)}\right\rangle=0 \quad \text{and} \quad \left\langle \xi_t^{(i)}\xi_{t'}^{(j)}\right\rangle = 2  k_B T  \gamma_i \delta_{ij}\delta(t-t').
\end{align}
The potential $U$ contains two distinct terms, it reads
\begin{align}
    &U
    ({\bf y},x)=  
        U_\mathrm{int}\left(y^{(2)}-y^{(1)}\right) + U_\mathrm{ext}\left(y^{(1)}-x\right),\label{eq:potential}
\end{align}
which encompasses an interaction potential $U_\mathrm{int}$ as well as a coupling of the probe to the protocol via an external potential $U_\mathrm{ext}$. We have in mind that this model mimics the case of a Brownian particle in a laser trap and interacting with a  viscoelastic surrounding. 

The interaction potential is nonlinear to allow for nonlinear effects, and it is specifically  a periodic potential~\cite{muller_properties_2020} 
\begin{align}
    U_\mathrm{int}(y) = - 2U_0 \cos \left( \frac{ \pi y}{d_1} \right)\sin \left( \frac{ \pi y}{d_2} \right)
    \label{eq:SPTPotential}
\end{align}
with amplitude $U_0$ and length scales $d_1$ and $d_2$. 
This interaction potential is non-binding, which allows, e.g., for pronounced nonlinear effects such as shear thinning~\cite{muller_properties_2020,jain_two_2021}. For $d_1=d_2$, we have $U_\mathrm{int}(y) = - U_0 \cos \left( \frac{2 \pi y}{d_1} -\frac{\pi}{2}\right)$, which is the symmetric potential used earlier~\cite{muller_properties_2020, caspers_nonlinear_2024}. For $d_1\not=d_2$, the potential is asymmetric. Such asymmetric case will be used to explore the identities involving even order responses. Even order responses vanish for the symmetric potential case.  

The external potential is chosen quadratic, mimicking, e.g., the potential of a harmonic trap,
\begin{align}
    U_\mathrm{ext}(y) = \frac{1}{2}\kappa y^2,
    \label{eq:trapPotential}
\end{align}
with stiffness $\kappa$.
The protocol $x_t$ thus presents the center of the trap, and changing its position as a function of time presents the perturbation. The force 
\begin{align}
F_t = \partial_{x_t} U({\bf y}_t,x_t)=-\kappa(y_t^{(1)}-x_t)\label{eq:forcetrap}
\end{align}
is proportional to the distance between probe and the center of the trap, which is measurable in typical experimental setups. 

We use the following parameter values. $\gamma_2 = 10\,\gamma_1$, i.e., particle 2 has a large friction compared to particle 1, yielding pronounced viscoelastic effects.
For the symmetric system $d_1=d_2 \equiv d$ and $U_0 = 2\, k_B T$, the latter giving rise to large nonlinear effects. For the asymmetric system $d_2 = 3\,d_1 = 1.5 \, d$ and $U_0 = k_B T$. For the overdamped case, $m_1=m_2=0$, while $m_1=m_2=\gamma_1^2 d^2/k_BT$ is used. The graphs are  expressed in terms of  $\gamma_1$, $d$, and $k_BT$.

\subsection{Testing identities, overdamped case}

\begin{figure}
    \centering
    \includegraphics{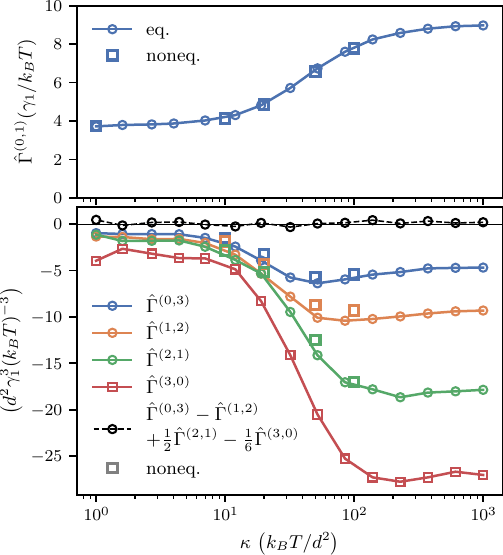}
    \caption{Kernels for the symmetric overdamped model with all time arguments integrated over, as a function of stiffness $\kappa$ of the harmonic trap. The top panel shows the first order kernels $\hat{\Gamma}^{(0,1)} = \hat{\Gamma}^{(1,0)}$ and the bottom panel shows the third order kernels. 
    Circles connected by lines are obtained from evaluation of the explicit forms for the kernels (Eqs.~(23)-(30) in Ref.~\cite{caspers_nonlinear_2024}),  squares are extracted from (non-)equilibrium force cumulants. $U_0 = 2\, k_B T$.}
    \label{fig:AllGamma_s=0}
\end{figure}

We start with the symmetric overdamped model, i.e., $d_1=d_2$ and $m_1=m_2=0$. In this case, the kernels are displayed in Fig.~\ref{fig:AllGamma_s=0}, as a function of the curvature $\kappa$ of the trapping potential. We evaluate the kernels with all time arguments integrated over, i.e., we evaluate them at $z=0$ for all arguments, with Laplace transform defined as $\hat{f}_z = \int_0^\infty \mathrm{d}t\, e^{-z t} f_t$. The kernels with $m+n=1$ are related in Eq.~\eqref{eq:id1}. This relation, being of linear order, resembles FDT, and it is displayed in Fig.~\ref{fig:AllGamma_s=0} for completeness. The identity for $m+n=2$, Eq.~\eqref{eq:id2}, has no content for the symmetric system, as all kernels vanish. Indeed, for reasons of symmetry, kernels with even sum   $m+n$  vanish.

The kernels with $m+n=3$ are finite and also shown in  
Fig.~\ref{fig:AllGamma_s=0}. In the graph, we show these kernels extracted in two ways. The first way is the "experimental" way, where we extract the force cumulants in simulations and from them the kernels (square symbols). The second way uses, as a double check, the explicit (microscopic) expressions for the kernels, which can be found for overdamped Brownian particles. These are given in Eqs.~(23)-(30) of Ref.~\cite{caspers_nonlinear_2024}, with the time-symmetric components explicitly given in Eqs.~(10) and (11) of Ref.~\cite{caspers_nonlinear_2024}. They thus require correlations to be taken in equilibrium simulations, and the results are given as circles in Fig.~\ref{fig:AllGamma_s=0}. 
The graph displays agreement between the methods, and also shows that the kernels indeed fulfill Eq.~\eqref{eq:id3}.      

\begin{figure}
    \centering
    \includegraphics{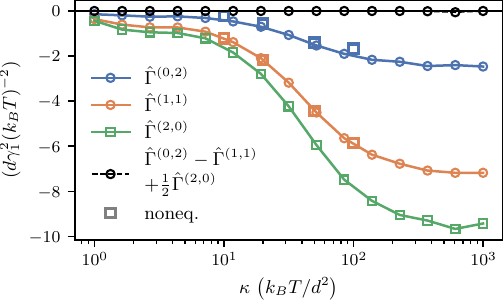}
    \caption{Second order kernels with all time arguments integrated over for the asymmetric overdamped model with $d_2 = 3\,d_1$ and $U_0 = k_B T$, for different stiffness $\kappa$. Circles with lines are obtained from direct evaluation kernel expressions (Eqs.~(24), (27) and (29) in Ref.~\cite{caspers_nonlinear_2024}), while squares are extracted from (non-)equilibrium force cumulants. }
    \label{fig:AllGamma2}
\end{figure}

Fig.~\ref{fig:AllGamma2} shows the kernels for an asymmetric potential, i.e, for $d_1\not=d_2$. In this case, also the kernels with 
$m+n=2$ are finite, and shown in the graphs. They are also evaluated via the two mentioned ways, using the kernel expressions, and extracted from the force cumulants. We find identity Eq.~\eqref{eq:id2} fulfilled. 

\subsection{Quantifying error of FDT form}
In this subsection we evaluate the identities in a different manner, namely by use of  Eqs.~\eqref{eq:cumulants} and \eqref{eq:DifferenceMeanForceconvolvedForceCov}. For these relations, the protocol must be specified, and we use a constant velocity protocol, i.e., $\dot x=v$.  For this case, Eq.~\eqref{eq:DifferenceMeanForceconvolvedForceCov} can be simplified to
\begin{align}
\begin{split}
    &\frac{\beta \langle F\rangle}{v} - \beta^2 \langle \hat{F}_{z=0};F_t\rangle\\
    &\quad= - \frac{v}{2} \hat{\Gamma}_{z=0,z=0}^{(2,0)}+ \frac{v^2}{6}  \hat{\Gamma}_{z=0,z=0,z=0}^{(3,0)} - \frac{v^2}{2} \hat{\Gamma}_{z=0;z=0,z=0}^{(2,1)}  \\
    &\quad\quad+ \mathcal{O}(v^3)\\
    &\quad= - \frac{\beta^3 v}{2} 
    \langle \hat{F}_{z=0};\hat{F}_{z=0};F_t\rangle
    + \frac{\beta^4 v^2}{6}  
     \langle \hat{F}_{z=0};\hat{F}_{z=0};\hat{F}_{z=0};F_t\rangle \\
    &\quad\quad+ \mathcal{O}(v^3) .
\end{split}\label{eq:DifferenceFlowCurveForceCov}
\end{align}
The first equality in Eq.~\eqref{eq:DifferenceFlowCurveForceCov} thereby is expressed in terms of the kernels, while the second equality is expressed in terms of force cumulants, providing two possibilities for determination of the right hand side in practice.  

We recall that the left hand side of Eq.~\eqref{eq:DifferenceFlowCurveForceCov} vanishes  in linear response, by virtue of the fluctuation dissipation theorem. The right hand side of Eq.~\eqref{eq:DifferenceMeanForceconvolvedForceCov} thus quantifies the error of FDT in higher orders. 

\begin{figure}
    \centering
    \includegraphics{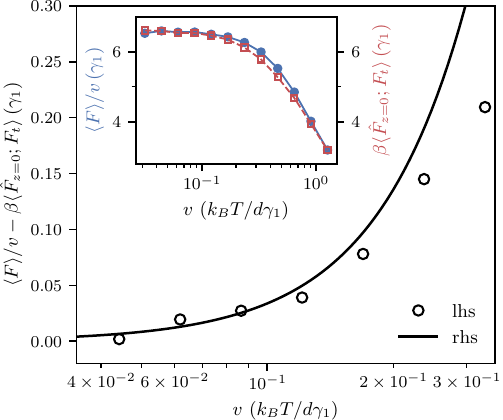}
    \caption{\textit{Main graph:}  Difference of mean force and force covariance for the symmetric overdamped model (symbols), together with the prediction to second order given by Eq.~\eqref{eq:DifferenceFlowCurveForceCov} (line).
    \textit{Inset:} Mean force (blue) and force covariance (red) for a large range of driving velocities $v$. $U_0 = 2\,k_B T$ and $\kappa = 50\, k_B T/d_1^2$.
    }
    \label{fig:NonlinearFDTCheckSPT}
\end{figure}

We start with the symmetric case, for which the right hand side of Eq.~\eqref{eq:DifferenceMeanForceconvolvedForceCov} is of order $v^2$. The inset of Fig.~\ref{fig:NonlinearFDTCheckSPT} shows the two terms on the lhs of Eq.~\eqref{eq:DifferenceFlowCurveForceCov}, i.e., mean force and force covariance as a function of protocol velocity $v$. For small $v$, the two agree, as mentioned, by reason of the fluctuation dissipation theorem. For larger $v$, they deviate. The deviation is shown in the main graph, together with the rhs of Eq.~\eqref{eq:DifferenceFlowCurveForceCov}. This difference is indeed of order $v^2$, and it agrees with  Eq.~\eqref{eq:DifferenceFlowCurveForceCov}. Notably, the deviation between the terms on the lhs is quite small, for all velocities shown. This might be coincidental.

\begin{figure}
    \centering
    \includegraphics{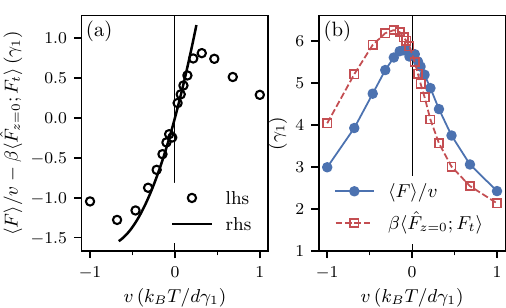}
    \caption{(a) Difference of mean force and force covariance for the asymmetric overdamped model with $d_2 = 3\, d_1$ (symbols), with the predictions to second order given by Eq.~\eqref{eq:DifferenceFlowCurveForceCov}. (b) Mean force (blue) and force covariance (red) for positive and negative driving velocities $v$. $U_0 = k_B T$ and $\kappa = 100\, k_B T/d_1^2$.}
    \label{fig:ExtendedFDTAsymmetricSPT}
\end{figure}

Fig.~\ref{fig:ExtendedFDTAsymmetricSPT} shows the asymmetric case, i.e, $d_1\not=d_2$, where the right hand side of Eq.~\eqref{eq:DifferenceFlowCurveForceCov} is of order $v$. Fig.~\ref{fig:ExtendedFDTAsymmetricSPT}(b) 
shows  the two terms on the lhs of Eq.~\ref{eq:DifferenceFlowCurveForceCov}, i.e., mean force and force covariance as a function of protocol velocity $v$. Here we extend to negative $v$, as the asymmetric model shows an asymmetric response curve. Both terms on the lhs of Eq.~\ref{eq:DifferenceFlowCurveForceCov} thus show asymmetric behavior as a function of $v$, and as expected, they agree for small $v$, and deviate for larger $v$.  The deviation is shown in panel (a), together with the rhs of Eq.~\eqref{eq:DifferenceFlowCurveForceCov}. We show the sum of first and second order terms of the rhs of Eq.~\eqref{eq:DifferenceFlowCurveForceCov}, which yield the slope and the curvature at the origin, with good agreement with the data points.

It is interesting to note that the response $\langle F\rangle$ is almost symmetric, while the covariance, i.e., the fluctuations, are strongly asymmetric.   

\subsection{Underdamped case}

\begin{figure}
    \centering
    \includegraphics{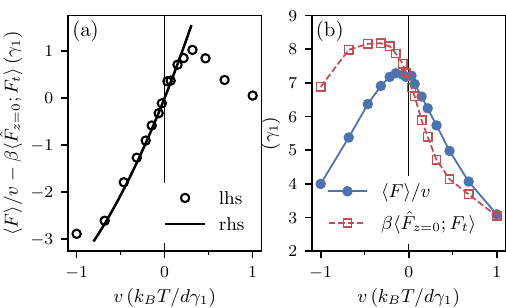}
    \caption{(a) Difference of mean force and force covariance for the asymmetric underdamped model with $d_2 = 3\,d_1$ and $m_1=m_2 =1$, with the predictions to second order given by Eq.~\eqref{eq:DifferenceFlowCurveForceCov}. (b) Mean force (blue) and force covariance (red) for positive and negative driving velocities $v$. $U_0 = k_B T$ and $\kappa = 100\, k_B T/d_1^2$.}
    \label{fig:ExtendedFDTUnderdampedAsymmetricSPT}
\end{figure}

As a final illustration and test, we employ the underdamped version of the asymmetric model, i.e., we use finite particle masses $m_1=m_2=\gamma_1^2 d^2/k_BT$. Fig.~\ref{fig:ExtendedFDTUnderdampedAsymmetricSPT}(a) shows the deviation of mean force and force covariance as a function of protocol velocity $v$. The rhs of Eq.~\eqref{eq:DifferenceFlowCurveForceCov} is shown up to second order and is extracted using the second identity in Eq.~\eqref{eq:DifferenceFlowCurveForceCov}. 
It shows good agreement with the slope and curvature of the data points at the origin.
Fig.~\ref{fig:ExtendedFDTUnderdampedAsymmetricSPT}(b) shows the two terms on the lhs of Eq.~\eqref{eq:DifferenceFlowCurveForceCov} with clear asymmetry and expected agreement for small $v$. 

Comparing Figs.~\ref{fig:ExtendedFDTAsymmetricSPT} and \ref{fig:ExtendedFDTUnderdampedAsymmetricSPT} displays that a finite mass seems to enhance the observed effects. Both the mean response as well as the variance are larger absolutely, and the asymmetry is also more pronounced. This hints that underdamped modes are especially sensitive to asymmetry.

\section{Conclusion}

We derived identities for non-equilibrium cumulants -- specifically for their memory kernels -- and tested them in various ways. These identities rely on local detailed balance, and can otherwise be applied to perturbations of the form $U({\bf y}_s, x_s)$, with a protocol $x_s$. The identities range from micro- to macroscopic scales. 

Future work can investigate these identities in higher dimensions, e.g., using multidimensional protocols. Types of protocols other than the examples investigated here are also of interest, for example time dependent particle interactions, or systems compressed in pistons.

Another task for future work concerns the derivation of identities for higher orders. While we have not checked these, the forms of Eqs.~\eqref{eq:id1} - \eqref{eq:id3} show a pattern that suggests a form of identity of order $n$,
\begin{align}
\begin{split}
  &  \Gamma_{s_1,\dots,s_n}^{(0,n)}\\&= \frac{1}{n!} \sum_{\pi \in S_n} \sum_{m=1}^n \frac{(-1)^{m+1}}{m!} 
    \Gamma_{s_{\pi(1)},\dots,s_{\pi(n-m)};s_{\pi(n-m+1)},\dots,s_{\pi(n)}}^{(m,n-m)},
\end{split}\label{eq:Gamma0nGuess}
\end{align}
compare also Ref.~\cite{holsten_thermodynamic_2021}.

\acknowledgements
This project was funded by the Deutsche Forschungsgemeinschaft (DFG), Grant No. SFB 1432 (Project ID 425217212) - Project C05. We thank Clemens Bechinger and Karthika Krishna Kumar for discussions. 

\appendix

\begin{widetext}
\section{Memory kernels}\label{sec:memory}

Recalling that all averages $\langle \cdot \rangle_\mathrm{eq}$ are evaluated in the equilibrium ensemble with protocol value $x_{t}$, the memory kernels entering the mean ($m=0$) read
\begin{align}
    \Gamma^{(0,0)} &= \beta \langle B \rangle_\mathrm{eq} \label{eq:Gamma00},\\
    \Gamma_{s}^{(0,1)} &= \beta^2 \langle F_s;B_0\rangle_\mathrm{eq} ,\label{eq:Gamma01g}\\
    \Gamma_{s_1,s_2}^{(0,2)} &= \frac{\beta^2}{2} \sum_{\pi \in S_2} \left( \left\langle \mathcal{D}_{s_{\pi(1)}};F_{s_{\pi(2)}};B_0\right\rangle_\mathrm{eq}+\left\langle F'_{s_{\pi(1)}};B_0\right\rangle_\mathrm{eq} \Theta\left(s_{\pi(1)}-s_{\pi(2)}\right) \right) ,\\
    \begin{split}
    \Gamma_{s_1,s_2,s_3}^{(0,3)} &= \frac{1}{6} \sum_{\pi \in S_3} \Bigg( \frac{\beta^2}{2} \left[\left\langle \mathcal{D}_{s_{\pi(1)}};\mathcal{D}_{s_{\pi(2)}};F_{s_{\pi(3)}};B_0\right\rangle_\mathrm{eq} - \left\langle \mathcal{D}_{s_{\pi(1)},s_{\pi(2)}};F_{s_{\pi(3)}};B_0\right\rangle_\mathrm{eq} \right] + \frac{\beta^4}{24} \left\langle F_{s_{\pi(1)}};F_{s_{\pi(2)}};F_{s_{\pi(3)}};B_0\right\rangle_\mathrm{eq} \\
    &\quad\quad\quad\quad\quad + \beta^2 \langle F_{s_{\pi(1)}}'';B_0\rangle_\mathrm{eq} \Theta\left(s_{\pi(1)}-s_{\pi(2)} \right)\Theta\left(s_{\pi(1)}-s_{\pi(3)} \right) + \beta^2\left\langle \mathcal{D}_{s_{\pi(1)}};F'_{s_{\pi(2)}};B_0 \right\rangle_\mathrm{eq} \Theta\left(s_{\pi(2)}-s_{\pi(3)}\right) \Bigg).
    \end{split} \\
    &\;\,\vdots \nonumber
\end{align}
In these expressions, the state observables $B$, $F$, $F'$ and so on are evaluated with the explicit dependence on $x_s$ set to $x_t$. 
Note that $\Gamma^{(0,0)}$ does not depend on the time instance of $\mathbf{y}$ which is why we omit the time index of $B$.
The sums in $\Gamma^{(0,2)}$ and $\Gamma^{(0,3)}$ go over all elements of the permutation group $S_2$ and $S_3$, respectively.
The first-order kernels entering the covariance ($m=1$) read
\begin{align}
    \Gamma_{t_1}^{(1,0)} &= \beta^2 \langle F_{t_1};B_0\rangle_\mathrm{eq}, \\
    \Gamma_{s;t_1}^{(1,1)} &= \beta^2 \left\langle \left(\mathcal{D}_s + \frac{\beta F_s}{2} \right);F_{t_1};B_0 \right\rangle_\mathrm{eq} + \beta^2 \left\langle F'_{t_1};B_0 \right\rangle_\mathrm{eq}\Theta \left(t_1-s\right), \\
    \begin{split}
    \Gamma_{s_1,s_2;t_1}^{(1,2)} &= \frac{\beta^2}{2} \sum_{\pi \in S_2} \Bigg[ \frac{1}{2}\left\langle \mathcal{D}_{s_{\pi(1)}};\mathcal{D}_{s_{\pi(2)}};F_{t_1};B_0\right\rangle_\mathrm{eq} - \frac{1}{2} \left\langle \mathcal{D}_{s_{\pi(1)},s_{\pi(2)}};F_{t_1};B_0\right\rangle_\mathrm{eq} + \frac{\beta}{2} \left\langle \mathcal{D}_{s_{\pi(1)}};F_{s_{\pi(2)}};F_{t_1};B_0\right\rangle_\mathrm{eq}\\
         &\quad\quad\quad + \frac{\beta^2}{8} \left\langle F_{s_{\pi(1)}};F_{s_{\pi(2)}};F_{t_1};B_0\right\rangle_\mathrm{eq} + \left(\frac{\beta}{2} \left\langle F_{s_{\pi(1)}};F'_{t_1};B_0 \right\rangle_\mathrm{eq} + \left\langle \mathcal{D}_{s_{\pi(1)}};F'_{t_1};B_0 \right\rangle_\mathrm{eq}\right)\Theta\left(t_1-s_{\pi(2)} \right)  \\
         &\quad\quad\quad + \frac{1}{2} \langle F''_{t_1};B_0\rangle_\mathrm{eq} \Theta\left(t_1-s_{\pi(1)}\right)\Theta\left(t_1-s_{\pi(2)}\right) + \frac{\beta}{2} \left\langle F'_{s_{\pi(1)}};F_{t_1};B_0 \right\rangle_\mathrm{eq} \Theta\left(s_{\pi(1)}-s_{\pi(2)}\right) 
        \Bigg].
    \end{split}\\
    &\;\;\vdots \nonumber
\end{align}
The kernels entering the third cumulant ($m=2$) are given by
\begin{align}
    \Gamma_{t_1,t_2}^{(2,0)} &= \beta^3 \langle F_{t_1};F_{t_2};B_0\rangle_\mathrm{eq} ,\\
    \begin{split}
    \Gamma_{s;t_1,t_2}^{(2,1)}  &=  \beta^3\left\langle \left( \mathcal{D}_{s}+\frac{\beta F_{s}}{2}  \right) ;F_{t_1};F_{t_2};B_0\right\rangle_\mathrm{eq} + \beta^3 \langle F'_{t_1};F_{t_2};B_0\rangle_\mathrm{eq}\Theta(t_1-s) \\
    &\quad + \beta^3 \langle F_{t_1};F'_{t_2};B_0\rangle_\mathrm{eq} \Theta(t_2-s),
    \end{split}\\
        &\;\;\vdots  \nonumber
\end{align}
and the zeroth-order kernel of the fourth cumulant ($m=3$) reads
\begin{align}
    \Gamma_{t_1,t_2,t_3}^{(3,0)} &= \beta^4 \langle F_{t_1};F_{t_2};F_{t_3};B_0\rangle_\mathrm{eq}.\label{eq:Gamma30g} \\
    &\;\;\vdots \nonumber
\end{align}
\end{widetext}

\section{Example: entropic component for underdamped Langevin dynamics}\label{app:EntropicComponentGeneralLangevin}

We consider a setup of $N$ underdamped Brownian degrees of freedom $\mathbf{y}$ with components $y^{(i)}$  ($i=1,\dots,N$) which are subject to a potential $U(\mathbf{y}_t,x_t)$.
$x_t$ is an external control parameter which couples to the Brownian degrees of freedom.
The setup is similar to that used in Ref.~\cite{caspers_nonlinear_2024}, but we consider underdamped dynamics and do not assume that the potential is translation invariant.
The corresponding Langevin equation for the degree of freedom $i$ is
\begin{align}
     m_i \ddot{y}_t^{(i)} = - \gamma_i \dot{y}_t^{(i)} -\partial_{y_t^{(i)}} U(\mathbf{y}_t,x_t) + \xi_t^{(i)},
     \label{eq:UnderdampedLangevin_y}
\end{align}
with Gaussian white noise $\xi_t^{(i)}$,
\begin{align}
    \left\langle \xi_t^{(i)}\right\rangle=0 \quad \text{and} \quad \left\langle \xi_t^{(i)}\xi_{t'}^{(j)}\right\rangle = 2  k_B T  \gamma_i \delta_{ij}\delta(t-t').
\end{align}
$m_i$ and $\gamma_i$ denote the mass and the bare friction coefficient of the degree of freedom $i$, respectively.
The system is prepared in equilibrium at time $t=t_0$ with the protocol being fixed at $x_{t_0}$.
Under the protocol $\smallx = (x_s,t_0 \leq s \leq t)$ the system evolves to non-equilibrium, which is characterized by a perturbation action $\mathcal{A}$, compare Eq.~\eqref{eq:PathintegralProtocolDependent}.
It is given by the difference of the Onsager-Machlup action functional~\cite{onsager_fluctuations_1953} for Eq.~\eqref{eq:UnderdampedLangevin_y} for $x_t$ and $x_t=x_{t_0}$,
\begin{align}
\begin{split}
    \mathcal{A} = \sum_i \frac{\gamma_i}{4 k_B T} \int_{t_0}^t \hspace{-0.2cm} \mathrm{d}s\, \Bigg[\left( \dot{y}_s^{(i)} + \frac{m_i}{\gamma_i} \ddot{y}_s^{(i)} +\frac{\partial_{y_s^{(i)}}U(\mathbf{y}_s,x_s)}{\gamma_i}  \right)^2 &\\
    - \left( \dot{y}_s^{(i)} + \frac{m_i}{\gamma_i} \ddot{y}_s^{(i)} +\frac{\partial_{y_s^{(i)}}U(\mathbf{y}_s,x_{t_0})}{\gamma_i}   \right)^2 \Bigg].&
    \end{split}
    \label{eq:PerturbationActionBrownianY}
\end{align}
Eq.~\eqref{eq:PerturbationActionBrownianY} simplifies to
\begin{align}
\begin{split}
    \mathcal{A} &= \sum_i \frac{1}{4 k_B T} \int_{t_0}^t \mathrm{d}s\, \Bigg\{ 2 \left(\dot{y}_s^{(i)}+ \frac{m_i \ddot{y}_s^{(i)}}{\gamma_i}\right)\\
    &\quad\quad \times\left[ \partial_{y_s^{(i)}}U(\mathbf{y}_s,x_{s})- \partial_{y_s^{(i)}}U(\mathbf{y}_s,x_{t_0})\right]\\
    &\quad+ \frac{1}{\gamma_i} \left[\left(\partial_{y_s^{(i)}}U(\mathbf{y}_s,x_{s}) \right)^2 -\left(\partial_{y_s^{(i)}}U(\mathbf{y}_s,x_{t_0}) \right)^2 \right] \Bigg\}.
    \end{split} 
\end{align}
Using the total derivative of the potential,
\begin{align}
    \frac{\mathrm{d}}{\mathrm{d}t} U(\mathbf{y}_t,x_t) = \sum_i \dot{y}_t^{(i)} \frac{\partial U(\mathbf{y}_t,x_t)}{\partial y_t^{(i)}} + \dot{x}_t F(\mathbf{y}_t,x_t),
\end{align}
with the conjugate observable $F(\mathbf{y}_t,x_t) = \partial_{x_t}U(\mathbf{y}_t,x_t)$,
we finally obtain for the action,
\begin{align}
\begin{split}
    \mathcal{A} &= \frac{1}{2 k_B T}  \left[ U(\mathbf{y}_t,x_t)  -  U(\mathbf{y}_t,x_{t_0})  -  \int_{t_0}^t \mathrm{d}s\, \dot{x}_s F(\mathbf{y}_s,x_s)  \right] \\
    &\quad + \sum_i \frac{m_i}{2 \gamma_i k_b T} \int_{t_0}^t \mathrm{d}s\, \ddot{y}_s^{(i)} \Big[ \partial_{y_s^{(i)}} U(\mathbf{y}_s,x_s) \\
    &\quad- \partial_{y_s^{(i)}} U(\mathbf{y}_s,x_{t_0}) \Big] +\sum_i \frac{1}{4 \gamma_i k_B T} \int_{t_0}^t \mathrm{d}s\,\\
    &\quad \times\left[\left(\partial_{y_s^{(i)}}U(\mathbf{y}_s,x_{s}) \right)^2 -\left(\partial_{y_s^{(i)}}U(\mathbf{y}_s,x_{t_0}) \right)^2 \right].
    \end{split} \label{eq:ActionUnderdampedLangevinFinal}
\end{align}
The last three lines are symmetric under time reversal; they form the non-thermodynamic component $D$ of the action.
The first line of Eq.~\eqref{eq:ActionUnderdampedLangevinFinal} is anti-symmetric under time reversal and gives the entropic component $S$.
Applying the definition $\mathcal{A} = D - S/2$, we find for the entropic component,
\begin{align}
    S =  \frac{1}{ k_B T}  \left[ U(\mathbf{y}_t,x_{t_0})  -  U(\mathbf{y}_t,x_{t})  + \int_{t_0}^t \mathrm{d}s\, \dot{x}_s F(\mathbf{y}_s,x_s) \right].\label{eq:SLangevinUprot}
\end{align}
The result is consistent with Eq.~\eqref{eq:SHamiltonianAnsatz} derived from thermodynamic considerations.

\bibliographystyle{apsrev4-2} 

\begin{thebibliography}{66}%
\makeatletter
\providecommand \@ifxundefined [1]{%
 \@ifx{#1\undefined}
}%
\providecommand \@ifnum [1]{%
 \ifnum #1\expandafter \@firstoftwo
 \else \expandafter \@secondoftwo
 \fi
}%
\providecommand \@ifx [1]{%
 \ifx #1\expandafter \@firstoftwo
 \else \expandafter \@secondoftwo
 \fi
}%
\providecommand \natexlab [1]{#1}%
\providecommand \enquote  [1]{``#1''}%
\providecommand \bibnamefont  [1]{#1}%
\providecommand \bibfnamefont [1]{#1}%
\providecommand \citenamefont [1]{#1}%
\providecommand \href@noop [0]{\@secondoftwo}%
\providecommand \href [0]{\begingroup \@sanitize@url \@href}%
\providecommand \@href[1]{\@@startlink{#1}\@@href}%
\providecommand \@@href[1]{\endgroup#1\@@endlink}%
\providecommand \@sanitize@url [0]{\catcode `\\12\catcode `\$12\catcode
  `\&12\catcode `\#12\catcode `\^12\catcode `\_12\catcode `\%12\relax}%
\providecommand \@@startlink[1]{}%
\providecommand \@@endlink[0]{}%
\providecommand \url  [0]{\begingroup\@sanitize@url \@url }%
\providecommand \@url [1]{\endgroup\@href {#1}{\urlprefix }}%
\providecommand \urlprefix  [0]{URL }%
\providecommand \Eprint [0]{\href }%
\providecommand \doibase [0]{https://doi.org/}%
\providecommand \selectlanguage [0]{\@gobble}%
\providecommand \bibinfo  [0]{\@secondoftwo}%
\providecommand \bibfield  [0]{\@secondoftwo}%
\providecommand \translation [1]{[#1]}%
\providecommand \BibitemOpen [0]{}%
\providecommand \bibitemStop [0]{}%
\providecommand \bibitemNoStop [0]{.\EOS\space}%
\providecommand \EOS [0]{\spacefactor3000\relax}%
\providecommand \BibitemShut  [1]{\csname bibitem#1\endcsname}%
\let\auto@bib@innerbib\@empty
\bibitem [{\citenamefont {Onsager}(1931)}]{onsager_reciprocal_1931}%
  \BibitemOpen
  \bibfield  {author} {\bibinfo {author} {\bibfnamefont {L.}~\bibnamefont
  {Onsager}},\ }\href {https://doi.org/10.1103/PhysRev.37.405} {\bibfield
  {journal} {\bibinfo  {journal} {Phys. Rev.}\ }\textbf {\bibinfo {volume}
  {37}},\ \bibinfo {pages} {405} (\bibinfo {year} {1931})}\BibitemShut
  {NoStop}%
\bibitem [{\citenamefont {Kawasaki}(1968)}]{kawasaki_theory_1968}%
  \BibitemOpen
  \bibfield  {author} {\bibinfo {author} {\bibfnamefont {T.}~\bibnamefont
  {Kawasaki}},\ }\href {https://doi.org/10.1143/PTP.39.331} {\bibfield
  {journal} {\bibinfo  {journal} {Prog. Theor. Phys.}\ }\textbf {\bibinfo
  {volume} {39}},\ \bibinfo {pages} {331} (\bibinfo {year} {1968})}\BibitemShut
  {NoStop}%
\bibitem [{\citenamefont {Evans}\ and\ \citenamefont
  {Morriss}(2008)}]{evans_statistical_2008}%
  \BibitemOpen
  \bibfield  {author} {\bibinfo {author} {\bibfnamefont {D.~J.}\ \bibnamefont
  {Evans}}\ and\ \bibinfo {author} {\bibfnamefont {G.}~\bibnamefont
  {Morriss}},\ }\href {https://doi.org/10.1017/CBO9780511535307} {\emph
  {\bibinfo {title} {Statistical {Mechanics} of {Nonequilibrium} {Liquids}}}},\
  \bibinfo {edition} {2nd}\ ed.\ (\bibinfo  {publisher} {Cambridge University
  Press},\ \bibinfo {year} {2008})\BibitemShut {NoStop}%
\bibitem [{\citenamefont {Fuchs}\ and\ \citenamefont
  {Cates}(2005)}]{fuchs_integration_2005}%
  \BibitemOpen
  \bibfield  {author} {\bibinfo {author} {\bibfnamefont {M.}~\bibnamefont
  {Fuchs}}\ and\ \bibinfo {author} {\bibfnamefont {M.~E.}\ \bibnamefont
  {Cates}},\ }\href {https://doi.org/10.1088/0953-8984/17/20/003} {\bibfield
  {journal} {\bibinfo  {journal} {J. Phys.: Condens. Matter}\ }\textbf
  {\bibinfo {volume} {17}},\ \bibinfo {pages} {S1681} (\bibinfo {year}
  {2005})}\BibitemShut {NoStop}%
\bibitem [{\citenamefont {Dhont}(1996)}]{dhont_introduction_1996}%
  \BibitemOpen
  \bibfield  {author} {\bibinfo {author} {\bibfnamefont {J.}~\bibnamefont
  {Dhont}},\ }\href {https://books.google.at/books?id=mmArTF5SJ9oC} {\emph
  {\bibinfo {title} {An {Introduction} to {Dynamics} of {Colloids}}}}\
  (\bibinfo  {publisher} {Elsevier Science},\ \bibinfo {year}
  {1996})\BibitemShut {NoStop}%
\bibitem [{\citenamefont {Zwanzig}(1961)}]{zwanzig_memory_1961}%
  \BibitemOpen
  \bibfield  {author} {\bibinfo {author} {\bibfnamefont {R.}~\bibnamefont
  {Zwanzig}},\ }\href {https://doi.org/10.1103/PhysRev.124.983} {\bibfield
  {journal} {\bibinfo  {journal} {Phys. Rev.}\ }\textbf {\bibinfo {volume}
  {124}},\ \bibinfo {pages} {983} (\bibinfo {year} {1961})}\BibitemShut
  {NoStop}%
\bibitem [{\citenamefont {Mori}(1965)}]{mori_transport_1965}%
  \BibitemOpen
  \bibfield  {author} {\bibinfo {author} {\bibfnamefont {H.}~\bibnamefont
  {Mori}},\ }\href {https://doi.org/10.1143/PTP.33.423} {\bibfield  {journal}
  {\bibinfo  {journal} {Prog. Theor. Phys.}\ }\textbf {\bibinfo {volume}
  {33}},\ \bibinfo {pages} {423} (\bibinfo {year} {1965})}\BibitemShut
  {NoStop}%
\bibitem [{\citenamefont {Zwanzig}(2001)}]{zwanzig_nonequilibrium_2001}%
  \BibitemOpen
  \bibfield  {author} {\bibinfo {author} {\bibfnamefont {R.}~\bibnamefont
  {Zwanzig}},\ }\href {https://books.google.de/books?id=4cI5136OdoMC} {\emph
  {\bibinfo {title} {Nonequilibrium {Statistical} {Mechanics}}}}\ (\bibinfo
  {publisher} {Oxford University Press},\ \bibinfo {year} {2001})\BibitemShut
  {NoStop}%
\bibitem [{\citenamefont {Grabert}(1982)}]{grabert_projection_1982}%
  \BibitemOpen
  \bibfield  {author} {\bibinfo {author} {\bibfnamefont {H.}~\bibnamefont
  {Grabert}},\ }\href {https://doi.org/10.1007/BFb0044591} {\emph {\bibinfo
  {title} {Projection {Operator} {Techniques} in {Nonequilibrium} {Statistical}
  {Mechanics}}}}\ (\bibinfo  {publisher} {Springer},\ \bibinfo {year}
  {1982})\BibitemShut {NoStop}%
\bibitem [{\citenamefont {Götze}(2008)}]{gotze_complex_2008}%
  \BibitemOpen
  \bibfield  {author} {\bibinfo {author} {\bibfnamefont {W.}~\bibnamefont
  {Götze}},\ }\href
  {https://doi.org/10.1093/acprof:oso/9780199235346.001.0001} {\emph {\bibinfo
  {title} {Complex {Dynamics} of {Glass}-{Forming} {Liquids}: {A}
  {Mode}-{Coupling} {Theory}}}}\ (\bibinfo  {publisher} {Oxford University
  Press},\ \bibinfo {year} {2008})\BibitemShut {NoStop}%
\bibitem [{\citenamefont {Janssen}(2018)}]{janssen_mode-coupling_2018}%
  \BibitemOpen
  \bibfield  {author} {\bibinfo {author} {\bibfnamefont {L.~M.~C.}\
  \bibnamefont {Janssen}},\ }\href
  {https://www.frontiersin.org/articles/10.3389/fphy.2018.00097} {\bibfield
  {journal} {\bibinfo  {journal} {Front. Phys.}\ }\textbf {\bibinfo {volume}
  {6}} (\bibinfo {year} {2018})}\BibitemShut {NoStop}%
\bibitem [{\citenamefont {Fuchs}\ \emph {et~al.}(1998)\citenamefont {Fuchs},
  \citenamefont {Götze},\ and\ \citenamefont {Mayr}}]{fuchs_asymptotic_1998}%
  \BibitemOpen
  \bibfield  {author} {\bibinfo {author} {\bibfnamefont {M.}~\bibnamefont
  {Fuchs}}, \bibinfo {author} {\bibfnamefont {W.}~\bibnamefont {Götze}},\ and\
  \bibinfo {author} {\bibfnamefont {M.~R.}\ \bibnamefont {Mayr}},\ }\href
  {https://doi.org/10.1103/PhysRevE.58.3384} {\bibfield  {journal} {\bibinfo
  {journal} {Phys. Rev. E}\ }\textbf {\bibinfo {volume} {58}},\ \bibinfo
  {pages} {3384} (\bibinfo {year} {1998})}\BibitemShut {NoStop}%
\bibitem [{\citenamefont {van Zanten}\ and\ \citenamefont
  {Rufener}(2000)}]{van_zanten_brownian_2000}%
  \BibitemOpen
  \bibfield  {author} {\bibinfo {author} {\bibfnamefont {J.~H.}\ \bibnamefont
  {van Zanten}}\ and\ \bibinfo {author} {\bibfnamefont {K.~P.}\ \bibnamefont
  {Rufener}},\ }\href {https://doi.org/10.1103/PhysRevE.62.5389} {\bibfield
  {journal} {\bibinfo  {journal} {Phys. Rev. E}\ }\textbf {\bibinfo {volume}
  {62}},\ \bibinfo {pages} {5389} (\bibinfo {year} {2000})}\BibitemShut
  {NoStop}%
\bibitem [{\citenamefont {Lu}\ and\ \citenamefont
  {Solomon}(2002)}]{lu_probe_2002}%
  \BibitemOpen
  \bibfield  {author} {\bibinfo {author} {\bibfnamefont {Q.}~\bibnamefont
  {Lu}}\ and\ \bibinfo {author} {\bibfnamefont {M.~J.}\ \bibnamefont
  {Solomon}},\ }\href {https://doi.org/10.1103/PhysRevE.66.061504} {\bibfield
  {journal} {\bibinfo  {journal} {Phys. Rev. E}\ }\textbf {\bibinfo {volume}
  {66}},\ \bibinfo {pages} {061504} (\bibinfo {year} {2002})}\BibitemShut
  {NoStop}%
\bibitem [{\citenamefont {Van Der~Gucht}\ \emph {et~al.}(2003)\citenamefont
  {Van Der~Gucht}, \citenamefont {Besseling}, \citenamefont {Knoben},
  \citenamefont {Bouteiller},\ and\ \citenamefont
  {Cohen~Stuart}}]{van_der_gucht_brownian_2003}%
  \BibitemOpen
  \bibfield  {author} {\bibinfo {author} {\bibfnamefont {J.}~\bibnamefont {Van
  Der~Gucht}}, \bibinfo {author} {\bibfnamefont {N.~A.~M.}\ \bibnamefont
  {Besseling}}, \bibinfo {author} {\bibfnamefont {W.}~\bibnamefont {Knoben}},
  \bibinfo {author} {\bibfnamefont {L.}~\bibnamefont {Bouteiller}},\ and\
  \bibinfo {author} {\bibfnamefont {M.~A.}\ \bibnamefont {Cohen~Stuart}},\
  }\href {https://doi.org/10.1103/PhysRevE.67.051106} {\bibfield  {journal}
  {\bibinfo  {journal} {Phys. Rev. E}\ }\textbf {\bibinfo {volume} {67}},\
  \bibinfo {pages} {051106} (\bibinfo {year} {2003})}\BibitemShut {NoStop}%
\bibitem [{\citenamefont {Caspers}\ \emph {et~al.}(2023)\citenamefont
  {Caspers}, \citenamefont {Ditz}, \citenamefont {Krishna~Kumar}, \citenamefont
  {Ginot}, \citenamefont {Bechinger}, \citenamefont {Fuchs},\ and\
  \citenamefont {Krüger}}]{caspers_how_2023}%
  \BibitemOpen
  \bibfield  {author} {\bibinfo {author} {\bibfnamefont {J.}~\bibnamefont
  {Caspers}}, \bibinfo {author} {\bibfnamefont {N.}~\bibnamefont {Ditz}},
  \bibinfo {author} {\bibfnamefont {K.}~\bibnamefont {Krishna~Kumar}}, \bibinfo
  {author} {\bibfnamefont {F.}~\bibnamefont {Ginot}}, \bibinfo {author}
  {\bibfnamefont {C.}~\bibnamefont {Bechinger}}, \bibinfo {author}
  {\bibfnamefont {M.}~\bibnamefont {Fuchs}},\ and\ \bibinfo {author}
  {\bibfnamefont {M.}~\bibnamefont {Krüger}},\ }\href
  {https://doi.org/10.1063/5.0129639} {\bibfield  {journal} {\bibinfo
  {journal} {J. Chem. Phys.}\ }\textbf {\bibinfo {volume} {158}},\ \bibinfo
  {pages} {024901} (\bibinfo {year} {2023})}\BibitemShut {NoStop}%
\bibitem [{\citenamefont {Caldeira}\ and\ \citenamefont
  {Leggett}(1981)}]{caldeira_influence_1981}%
  \BibitemOpen
  \bibfield  {author} {\bibinfo {author} {\bibfnamefont {A.~O.}\ \bibnamefont
  {Caldeira}}\ and\ \bibinfo {author} {\bibfnamefont {A.~J.}\ \bibnamefont
  {Leggett}},\ }\href {https://doi.org/10.1103/PhysRevLett.46.211} {\bibfield
  {journal} {\bibinfo  {journal} {Phys. Rev. Lett.}\ }\textbf {\bibinfo
  {volume} {46}},\ \bibinfo {pages} {211} (\bibinfo {year} {1981})}\BibitemShut
  {NoStop}%
\bibitem [{\citenamefont {Doerries}\ \emph {et~al.}(2021)\citenamefont
  {Doerries}, \citenamefont {Loos},\ and\ \citenamefont
  {Klapp}}]{doerries_correlation_2021}%
  \BibitemOpen
  \bibfield  {author} {\bibinfo {author} {\bibfnamefont {T.~J.}\ \bibnamefont
  {Doerries}}, \bibinfo {author} {\bibfnamefont {S.~A.~M.}\ \bibnamefont
  {Loos}},\ and\ \bibinfo {author} {\bibfnamefont {S.~H.~L.}\ \bibnamefont
  {Klapp}},\ }\href {https://doi.org/10.1088/1742-5468/abdead} {\bibfield
  {journal} {\bibinfo  {journal} {J. Stat. Mech.}\ }\textbf {\bibinfo {volume}
  {2021}},\ \bibinfo {pages} {033202} (\bibinfo {year} {2021})}\BibitemShut
  {NoStop}%
\bibitem [{\citenamefont {Venturelli}\ and\ \citenamefont
  {Gambassi}(2023)}]{venturelli_memory-induced_2023}%
  \BibitemOpen
  \bibfield  {author} {\bibinfo {author} {\bibfnamefont {D.}~\bibnamefont
  {Venturelli}}\ and\ \bibinfo {author} {\bibfnamefont {A.}~\bibnamefont
  {Gambassi}},\ }\href
  {https://iopscience.iop.org/article/10.1088/1367-2630/acf240/meta} {\bibfield
   {journal} {\bibinfo  {journal} {New J. Phys.}\ }\textbf {\bibinfo {volume}
  {25}},\ \bibinfo {pages} {093025} (\bibinfo {year} {2023})}\BibitemShut
  {NoStop}%
\bibitem [{\citenamefont {Squires}\ and\ \citenamefont
  {Brady}(2005)}]{squires_simple_2005}%
  \BibitemOpen
  \bibfield  {author} {\bibinfo {author} {\bibfnamefont {T.~M.}\ \bibnamefont
  {Squires}}\ and\ \bibinfo {author} {\bibfnamefont {J.~F.}\ \bibnamefont
  {Brady}},\ }\href {https://doi.org/10.1063/1.1960607} {\bibfield  {journal}
  {\bibinfo  {journal} {Phys. Fluids}\ }\textbf {\bibinfo {volume} {17}},\
  \bibinfo {pages} {073101} (\bibinfo {year} {2005})}\BibitemShut {NoStop}%
\bibitem [{\citenamefont {Gazuz}\ \emph {et~al.}(2009)\citenamefont {Gazuz},
  \citenamefont {Puertas}, \citenamefont {Voigtmann},\ and\ \citenamefont
  {Fuchs}}]{gazuz_active_2009}%
  \BibitemOpen
  \bibfield  {author} {\bibinfo {author} {\bibfnamefont {I.}~\bibnamefont
  {Gazuz}}, \bibinfo {author} {\bibfnamefont {A.~M.}\ \bibnamefont {Puertas}},
  \bibinfo {author} {\bibfnamefont {T.}~\bibnamefont {Voigtmann}},\ and\
  \bibinfo {author} {\bibfnamefont {M.}~\bibnamefont {Fuchs}},\ }\href
  {https://doi.org/10.1103/PhysRevLett.102.248302} {\bibfield  {journal}
  {\bibinfo  {journal} {Phys. Rev. Lett.}\ }\textbf {\bibinfo {volume} {102}},\
  \bibinfo {pages} {248302} (\bibinfo {year} {2009})}\BibitemShut {NoStop}%
\bibitem [{\citenamefont {Harrer}\ \emph {et~al.}(2012)\citenamefont {Harrer},
  \citenamefont {Winter}, \citenamefont {Horbach}, \citenamefont {Fuchs},\ and\
  \citenamefont {Voigtmann}}]{harrer_force-induced_2012}%
  \BibitemOpen
  \bibfield  {author} {\bibinfo {author} {\bibfnamefont {C.~J.}\ \bibnamefont
  {Harrer}}, \bibinfo {author} {\bibfnamefont {D.}~\bibnamefont {Winter}},
  \bibinfo {author} {\bibfnamefont {J.}~\bibnamefont {Horbach}}, \bibinfo
  {author} {\bibfnamefont {M.}~\bibnamefont {Fuchs}},\ and\ \bibinfo {author}
  {\bibfnamefont {T.}~\bibnamefont {Voigtmann}},\ }\href
  {https://doi.org/10.1088/0953-8984/24/46/464105} {\bibfield  {journal}
  {\bibinfo  {journal} {J. Phys.: Condens. Matter}\ }\textbf {\bibinfo {volume}
  {24}},\ \bibinfo {pages} {464105} (\bibinfo {year} {2012})}\BibitemShut
  {NoStop}%
\bibitem [{\citenamefont {Jayaraman}\ and\ \citenamefont
  {Belmonte}(2003)}]{jayaraman_oscillations_2003}%
  \BibitemOpen
  \bibfield  {author} {\bibinfo {author} {\bibfnamefont {A.}~\bibnamefont
  {Jayaraman}}\ and\ \bibinfo {author} {\bibfnamefont {A.}~\bibnamefont
  {Belmonte}},\ }\href {https://doi.org/10.1103/PhysRevE.67.065301} {\bibfield
  {journal} {\bibinfo  {journal} {Phys. Rev. E}\ }\textbf {\bibinfo {volume}
  {67}},\ \bibinfo {pages} {065301} (\bibinfo {year} {2003})}\BibitemShut
  {NoStop}%
\bibitem [{\citenamefont {Handzy}\ and\ \citenamefont
  {Belmonte}(2004)}]{handzy_oscillatory_2004}%
  \BibitemOpen
  \bibfield  {author} {\bibinfo {author} {\bibfnamefont {N.~Z.}\ \bibnamefont
  {Handzy}}\ and\ \bibinfo {author} {\bibfnamefont {A.}~\bibnamefont
  {Belmonte}},\ }\href {https://doi.org/10.1103/PhysRevLett.92.124501}
  {\bibfield  {journal} {\bibinfo  {journal} {Phys. Rev. Lett.}\ }\textbf
  {\bibinfo {volume} {92}},\ \bibinfo {pages} {124501} (\bibinfo {year}
  {2004})}\BibitemShut {NoStop}%
\bibitem [{\citenamefont {Berner}\ \emph {et~al.}(2018)\citenamefont {Berner},
  \citenamefont {Müller}, \citenamefont {Gomez-Solano}, \citenamefont
  {Krüger},\ and\ \citenamefont {Bechinger}}]{berner_oscillating_2018}%
  \BibitemOpen
  \bibfield  {author} {\bibinfo {author} {\bibfnamefont {J.}~\bibnamefont
  {Berner}}, \bibinfo {author} {\bibfnamefont {B.}~\bibnamefont {Müller}},
  \bibinfo {author} {\bibfnamefont {J.~R.}\ \bibnamefont {Gomez-Solano}},
  \bibinfo {author} {\bibfnamefont {M.}~\bibnamefont {Krüger}},\ and\ \bibinfo
  {author} {\bibfnamefont {C.}~\bibnamefont {Bechinger}},\ }\href
  {https://doi.org/10.1038/s41467-018-03345-2} {\bibfield  {journal} {\bibinfo
  {journal} {Nat. Commun.}\ }\textbf {\bibinfo {volume} {9}},\ \bibinfo {pages}
  {999} (\bibinfo {year} {2018})}\BibitemShut {NoStop}%
\bibitem [{\citenamefont {Jain}\ \emph {et~al.}(2021)\citenamefont {Jain},
  \citenamefont {Ginot}, \citenamefont {Berner}, \citenamefont {Bechinger},\
  and\ \citenamefont {Krüger}}]{jain_two_2021}%
  \BibitemOpen
  \bibfield  {author} {\bibinfo {author} {\bibfnamefont {R.}~\bibnamefont
  {Jain}}, \bibinfo {author} {\bibfnamefont {F.}~\bibnamefont {Ginot}},
  \bibinfo {author} {\bibfnamefont {J.}~\bibnamefont {Berner}}, \bibinfo
  {author} {\bibfnamefont {C.}~\bibnamefont {Bechinger}},\ and\ \bibinfo
  {author} {\bibfnamefont {M.}~\bibnamefont {Krüger}},\ }\href
  {https://doi.org/10.1063/5.0048320} {\bibfield  {journal} {\bibinfo
  {journal} {J. Chem. Phys.}\ }\textbf {\bibinfo {volume} {154}},\ \bibinfo
  {pages} {184904} (\bibinfo {year} {2021})}\BibitemShut {NoStop}%
\bibitem [{\citenamefont {Winter}\ \emph {et~al.}(2012)\citenamefont {Winter},
  \citenamefont {Horbach}, \citenamefont {Virnau},\ and\ \citenamefont
  {Binder}}]{winter_active_2012}%
  \BibitemOpen
  \bibfield  {author} {\bibinfo {author} {\bibfnamefont {D.}~\bibnamefont
  {Winter}}, \bibinfo {author} {\bibfnamefont {J.}~\bibnamefont {Horbach}},
  \bibinfo {author} {\bibfnamefont {P.}~\bibnamefont {Virnau}},\ and\ \bibinfo
  {author} {\bibfnamefont {K.}~\bibnamefont {Binder}},\ }\href
  {https://doi.org/10.1103/PhysRevLett.108.028303} {\bibfield  {journal}
  {\bibinfo  {journal} {Phys. Rev. Lett.}\ }\textbf {\bibinfo {volume} {108}},\
  \bibinfo {pages} {028303} (\bibinfo {year} {2012})}\BibitemShut {NoStop}%
\bibitem [{\citenamefont {Bénichou}\ \emph {et~al.}(2013)\citenamefont
  {Bénichou}, \citenamefont {Bodrova}, \citenamefont {Chakraborty},
  \citenamefont {Illien}, \citenamefont {Law}, \citenamefont
  {Mejía-Monasterio}, \citenamefont {Oshanin},\ and\ \citenamefont
  {Voituriez}}]{benichou_geometry-induced_2013}%
  \BibitemOpen
  \bibfield  {author} {\bibinfo {author} {\bibfnamefont {O.}~\bibnamefont
  {Bénichou}}, \bibinfo {author} {\bibfnamefont {A.}~\bibnamefont {Bodrova}},
  \bibinfo {author} {\bibfnamefont {D.}~\bibnamefont {Chakraborty}}, \bibinfo
  {author} {\bibfnamefont {P.}~\bibnamefont {Illien}}, \bibinfo {author}
  {\bibfnamefont {A.}~\bibnamefont {Law}}, \bibinfo {author} {\bibfnamefont
  {C.}~\bibnamefont {Mejía-Monasterio}}, \bibinfo {author} {\bibfnamefont
  {G.}~\bibnamefont {Oshanin}},\ and\ \bibinfo {author} {\bibfnamefont
  {R.}~\bibnamefont {Voituriez}},\ }\href
  {https://doi.org/10.1103/PhysRevLett.111.260601} {\bibfield  {journal}
  {\bibinfo  {journal} {Phys. Rev. Lett.}\ }\textbf {\bibinfo {volume} {111}},\
  \bibinfo {pages} {260601} (\bibinfo {year} {2013})}\BibitemShut {NoStop}%
\bibitem [{\citenamefont {Wilson}\ and\ \citenamefont
  {Poon}(2011)}]{wilson_small-world_2011}%
  \BibitemOpen
  \bibfield  {author} {\bibinfo {author} {\bibfnamefont {L.~G.}\ \bibnamefont
  {Wilson}}\ and\ \bibinfo {author} {\bibfnamefont {W.~C.~K.}\ \bibnamefont
  {Poon}},\ }\href {https://doi.org/10.1039/C0CP01564D} {\bibfield  {journal}
  {\bibinfo  {journal} {Phys. Chem. Chem. Phys.}\ }\textbf {\bibinfo {volume}
  {13}},\ \bibinfo {pages} {10617} (\bibinfo {year} {2011})}\BibitemShut
  {NoStop}%
\bibitem [{\citenamefont {Démery}\ and\ \citenamefont
  {Fodor}(2019)}]{demery_driven_2019}%
  \BibitemOpen
  \bibfield  {author} {\bibinfo {author} {\bibfnamefont {V.}~\bibnamefont
  {Démery}}\ and\ \bibinfo {author} {\bibfnamefont {E.}~\bibnamefont
  {Fodor}},\ }\href {https://doi.org/10.1088/1742-5468/ab02e9} {\bibfield
  {journal} {\bibinfo  {journal} {J. Stat. Mech.}\ }\textbf {\bibinfo {volume}
  {2019}},\ \bibinfo {pages} {033202} (\bibinfo {year} {2019})}\BibitemShut
  {NoStop}%
\bibitem [{\citenamefont {Cugliandolo}(2011)}]{cugliandolo_effective_2011}%
  \BibitemOpen
  \bibfield  {author} {\bibinfo {author} {\bibfnamefont {L.~F.}\ \bibnamefont
  {Cugliandolo}},\ }\href {https://doi.org/10.1088/1751-8113/44/48/483001}
  {\bibfield  {journal} {\bibinfo  {journal} {J. Phys. A: Math. Theor.}\
  }\textbf {\bibinfo {volume} {44}},\ \bibinfo {pages} {483001} (\bibinfo
  {year} {2011})}\BibitemShut {NoStop}%
\bibitem [{\citenamefont {Puglisi}\ \emph {et~al.}(2017)\citenamefont
  {Puglisi}, \citenamefont {Sarracino},\ and\ \citenamefont
  {Vulpiani}}]{puglisi_temperature_2017}%
  \BibitemOpen
  \bibfield  {author} {\bibinfo {author} {\bibfnamefont {A.}~\bibnamefont
  {Puglisi}}, \bibinfo {author} {\bibfnamefont {A.}~\bibnamefont {Sarracino}},\
  and\ \bibinfo {author} {\bibfnamefont {A.}~\bibnamefont {Vulpiani}},\ }\href
  {https://doi.org/10.1016/j.physrep.2017.09.001} {\bibfield  {journal}
  {\bibinfo  {journal} {Phys. Rep.}\ }\textbf {\bibinfo {volume} {709-710}},\
  \bibinfo {pages} {1} (\bibinfo {year} {2017})}\BibitemShut {NoStop}%
\bibitem [{\citenamefont {Brader}\ \emph {et~al.}(2008)\citenamefont {Brader},
  \citenamefont {Cates},\ and\ \citenamefont
  {Fuchs}}]{brader_first-principles_2008}%
  \BibitemOpen
  \bibfield  {author} {\bibinfo {author} {\bibfnamefont {J.~M.}\ \bibnamefont
  {Brader}}, \bibinfo {author} {\bibfnamefont {M.~E.}\ \bibnamefont {Cates}},\
  and\ \bibinfo {author} {\bibfnamefont {M.}~\bibnamefont {Fuchs}},\ }\href
  {https://doi.org/10.1103/PhysRevLett.101.138301} {\bibfield  {journal}
  {\bibinfo  {journal} {Phys. Rev. Lett.}\ }\textbf {\bibinfo {volume} {101}},\
  \bibinfo {pages} {138301} (\bibinfo {year} {2008})}\BibitemShut {NoStop}%
\bibitem [{\citenamefont {Gazuz}\ and\ \citenamefont
  {Fuchs}(2013)}]{gazuz_nonlinear_2013}%
  \BibitemOpen
  \bibfield  {author} {\bibinfo {author} {\bibfnamefont {I.}~\bibnamefont
  {Gazuz}}\ and\ \bibinfo {author} {\bibfnamefont {M.}~\bibnamefont {Fuchs}},\
  }\href {https://doi.org/10.1103/PhysRevE.87.032304} {\bibfield  {journal}
  {\bibinfo  {journal} {Phys. Rev. E}\ }\textbf {\bibinfo {volume} {87}},\
  \bibinfo {pages} {032304} (\bibinfo {year} {2013})}\BibitemShut {NoStop}%
\bibitem [{\citenamefont {Gruber}\ \emph {et~al.}(2016)\citenamefont {Gruber},
  \citenamefont {Abade}, \citenamefont {Puertas},\ and\ \citenamefont
  {Fuchs}}]{gruber_active_2016}%
  \BibitemOpen
  \bibfield  {author} {\bibinfo {author} {\bibfnamefont {M.}~\bibnamefont
  {Gruber}}, \bibinfo {author} {\bibfnamefont {G.~C.}\ \bibnamefont {Abade}},
  \bibinfo {author} {\bibfnamefont {A.~M.}\ \bibnamefont {Puertas}},\ and\
  \bibinfo {author} {\bibfnamefont {M.}~\bibnamefont {Fuchs}},\ }\href
  {https://doi.org/10.1103/PhysRevE.94.042602} {\bibfield  {journal} {\bibinfo
  {journal} {Phys. Rev. E}\ }\textbf {\bibinfo {volume} {94}},\ \bibinfo
  {pages} {042602} (\bibinfo {year} {2016})}\BibitemShut {NoStop}%
\bibitem [{\citenamefont {Meyer}\ \emph {et~al.}(2017)\citenamefont {Meyer},
  \citenamefont {Voigtmann},\ and\ \citenamefont
  {Schilling}}]{meyer_non-stationary_2017}%
  \BibitemOpen
  \bibfield  {author} {\bibinfo {author} {\bibfnamefont {H.}~\bibnamefont
  {Meyer}}, \bibinfo {author} {\bibfnamefont {T.}~\bibnamefont {Voigtmann}},\
  and\ \bibinfo {author} {\bibfnamefont {T.}~\bibnamefont {Schilling}},\ }\href
  {https://doi.org/10.1063/1.5006980} {\bibfield  {journal} {\bibinfo
  {journal} {J. Chem. Phys.}\ }\textbf {\bibinfo {volume} {147}},\ \bibinfo
  {pages} {214110} (\bibinfo {year} {2017})}\BibitemShut {NoStop}%
\bibitem [{\citenamefont {te~Vrugt}\ and\ \citenamefont
  {Wittkowski}(2019)}]{te_vrugt_mori-zwanzig_2019}%
  \BibitemOpen
  \bibfield  {author} {\bibinfo {author} {\bibfnamefont {M.}~\bibnamefont
  {te~Vrugt}}\ and\ \bibinfo {author} {\bibfnamefont {R.}~\bibnamefont
  {Wittkowski}},\ }\href {https://doi.org/10.1103/PhysRevE.99.062118}
  {\bibfield  {journal} {\bibinfo  {journal} {Phys. Rev. E}\ }\textbf {\bibinfo
  {volume} {99}},\ \bibinfo {pages} {062118} (\bibinfo {year}
  {2019})}\BibitemShut {NoStop}%
\bibitem [{\citenamefont {Meyer}\ \emph {et~al.}(2019)\citenamefont {Meyer},
  \citenamefont {Voigtmann},\ and\ \citenamefont
  {Schilling}}]{meyer_dynamics_2019}%
  \BibitemOpen
  \bibfield  {author} {\bibinfo {author} {\bibfnamefont {H.}~\bibnamefont
  {Meyer}}, \bibinfo {author} {\bibfnamefont {T.}~\bibnamefont {Voigtmann}},\
  and\ \bibinfo {author} {\bibfnamefont {T.}~\bibnamefont {Schilling}},\ }\href
  {https://doi.org/10.1063/1.5090450} {\bibfield  {journal} {\bibinfo
  {journal} {J. Chem. Phys.}\ }\textbf {\bibinfo {volume} {150}},\ \bibinfo
  {pages} {174118} (\bibinfo {year} {2019})}\BibitemShut {NoStop}%
\bibitem [{\citenamefont {Glatzel}\ and\ \citenamefont
  {Schilling}(2021)}]{glatzel_interplay_2021}%
  \BibitemOpen
  \bibfield  {author} {\bibinfo {author} {\bibfnamefont {F.}~\bibnamefont
  {Glatzel}}\ and\ \bibinfo {author} {\bibfnamefont {T.}~\bibnamefont
  {Schilling}},\ }\href {https://doi.org/10.1209/0295-5075/ac35ba} {\bibfield
  {journal} {\bibinfo  {journal} {EPL}\ }\textbf {\bibinfo {volume} {136}},\
  \bibinfo {pages} {36001} (\bibinfo {year} {2021})}\BibitemShut {NoStop}%
\bibitem [{\citenamefont {Vroylandt}\ and\ \citenamefont
  {Monmarché}(2022)}]{vroylandt_position-dependent_2022}%
  \BibitemOpen
  \bibfield  {author} {\bibinfo {author} {\bibfnamefont {H.}~\bibnamefont
  {Vroylandt}}\ and\ \bibinfo {author} {\bibfnamefont {P.}~\bibnamefont
  {Monmarché}},\ }\href {https://doi.org/10.1063/5.0094566} {\bibfield
  {journal} {\bibinfo  {journal} {J. Chem. Phys.}\ }\textbf {\bibinfo {volume}
  {156}},\ \bibinfo {pages} {244105} (\bibinfo {year} {2022})}\BibitemShut
  {NoStop}%
\bibitem [{\citenamefont {Schilling}(2022)}]{schilling_coarse-grained_2022}%
  \BibitemOpen
  \bibfield  {author} {\bibinfo {author} {\bibfnamefont {T.}~\bibnamefont
  {Schilling}},\ }\href {https://doi.org/10.1016/j.physrep.2022.04.006}
  {\bibfield  {journal} {\bibinfo  {journal} {Phys. Rep.}\ }\textbf {\bibinfo
  {volume} {972}},\ \bibinfo {pages} {1} (\bibinfo {year} {2022})}\BibitemShut
  {NoStop}%
\bibitem [{\citenamefont {Netz}(2024)}]{netz_derivation_2024}%
  \BibitemOpen
  \bibfield  {author} {\bibinfo {author} {\bibfnamefont {R.~R.}\ \bibnamefont
  {Netz}},\ }\href {https://doi.org/10.1103/PhysRevE.110.014123} {\bibfield
  {journal} {\bibinfo  {journal} {Phys. Rev. E}\ }\textbf {\bibinfo {volume}
  {110}},\ \bibinfo {pages} {014123} (\bibinfo {year} {2024})}\BibitemShut
  {NoStop}%
\bibitem [{\citenamefont {Penna}\ \emph {et~al.}(2003)\citenamefont {Penna},
  \citenamefont {Dzubiella},\ and\ \citenamefont
  {Tarazona}}]{penna_dynamic_2003}%
  \BibitemOpen
  \bibfield  {author} {\bibinfo {author} {\bibfnamefont {F.}~\bibnamefont
  {Penna}}, \bibinfo {author} {\bibfnamefont {J.}~\bibnamefont {Dzubiella}},\
  and\ \bibinfo {author} {\bibfnamefont {P.}~\bibnamefont {Tarazona}},\ }\href
  {https://doi.org/10.1103/PhysRevE.68.061407} {\bibfield  {journal} {\bibinfo
  {journal} {Phys. Rev. E}\ }\textbf {\bibinfo {volume} {68}},\ \bibinfo
  {pages} {061407} (\bibinfo {year} {2003})}\BibitemShut {NoStop}%
\bibitem [{\citenamefont {Rauscher}\ \emph {et~al.}(2007)\citenamefont
  {Rauscher}, \citenamefont {Domínguez}, \citenamefont {Krüger},\ and\
  \citenamefont {Penna}}]{rauscher_dynamic_2007}%
  \BibitemOpen
  \bibfield  {author} {\bibinfo {author} {\bibfnamefont {M.}~\bibnamefont
  {Rauscher}}, \bibinfo {author} {\bibfnamefont {A.}~\bibnamefont
  {Domínguez}}, \bibinfo {author} {\bibfnamefont {M.}~\bibnamefont
  {Krüger}},\ and\ \bibinfo {author} {\bibfnamefont {F.}~\bibnamefont
  {Penna}},\ }\href {https://doi.org/10.1063/1.2806094} {\bibfield  {journal}
  {\bibinfo  {journal} {J. Chem. Phys.}\ }\textbf {\bibinfo {volume} {127}},\
  \bibinfo {pages} {244906} (\bibinfo {year} {2007})}\BibitemShut {NoStop}%
\bibitem [{\citenamefont {De~Las~Heras}\ and\ \citenamefont
  {Schmidt}(2018)}]{de_las_heras_velocity_2018}%
  \BibitemOpen
  \bibfield  {author} {\bibinfo {author} {\bibfnamefont {D.}~\bibnamefont
  {De~Las~Heras}}\ and\ \bibinfo {author} {\bibfnamefont {M.}~\bibnamefont
  {Schmidt}},\ }\href {https://doi.org/10.1103/PhysRevLett.120.028001}
  {\bibfield  {journal} {\bibinfo  {journal} {Phys. Rev. Lett.}\ }\textbf
  {\bibinfo {volume} {120}},\ \bibinfo {pages} {028001} (\bibinfo {year}
  {2018})}\BibitemShut {NoStop}%
\bibitem [{\citenamefont {Schmidt}(2022)}]{schmidt_power_2022}%
  \BibitemOpen
  \bibfield  {author} {\bibinfo {author} {\bibfnamefont {M.}~\bibnamefont
  {Schmidt}},\ }\href {https://doi.org/10.1103/RevModPhys.94.015007} {\bibfield
   {journal} {\bibinfo  {journal} {Rev. Mod. Phys.}\ }\textbf {\bibinfo
  {volume} {94}},\ \bibinfo {pages} {015007} (\bibinfo {year}
  {2022})}\BibitemShut {NoStop}%
\bibitem [{\citenamefont {Leitmann}\ and\ \citenamefont
  {Franosch}(2013)}]{leitmann_nonlinear_2013}%
  \BibitemOpen
  \bibfield  {author} {\bibinfo {author} {\bibfnamefont {S.}~\bibnamefont
  {Leitmann}}\ and\ \bibinfo {author} {\bibfnamefont {T.}~\bibnamefont
  {Franosch}},\ }\href {https://doi.org/10.1103/PhysRevLett.111.190603}
  {\bibfield  {journal} {\bibinfo  {journal} {Phys. Rev. Lett.}\ }\textbf
  {\bibinfo {volume} {111}},\ \bibinfo {pages} {190603} (\bibinfo {year}
  {2013})}\BibitemShut {NoStop}%
\bibitem [{\citenamefont {Leitmann}\ \emph {et~al.}(2018)\citenamefont
  {Leitmann}, \citenamefont {Bénichou},\ and\ \citenamefont
  {Franosch}}]{leitmann_time-dependent_2018}%
  \BibitemOpen
  \bibfield  {author} {\bibinfo {author} {\bibfnamefont {S.}~\bibnamefont
  {Leitmann}}, \bibinfo {author} {\bibfnamefont {O.}~\bibnamefont
  {Bénichou}},\ and\ \bibinfo {author} {\bibfnamefont {T.}~\bibnamefont
  {Franosch}},\ }\href {https://doi.org/10.1088/1751-8121/aad341} {\bibfield
  {journal} {\bibinfo  {journal} {J. Phys. A: Math. Theor.}\ }\textbf {\bibinfo
  {volume} {51}},\ \bibinfo {pages} {375001} (\bibinfo {year}
  {2018})}\BibitemShut {NoStop}%
\bibitem [{\citenamefont {Asheichyk}\ \emph {et~al.}(2021)\citenamefont
  {Asheichyk}, \citenamefont {Fuchs},\ and\ \citenamefont
  {Krüger}}]{asheichyk_brownian_2021}%
  \BibitemOpen
  \bibfield  {author} {\bibinfo {author} {\bibfnamefont {K.}~\bibnamefont
  {Asheichyk}}, \bibinfo {author} {\bibfnamefont {M.}~\bibnamefont {Fuchs}},\
  and\ \bibinfo {author} {\bibfnamefont {M.}~\bibnamefont {Krüger}},\ }\href
  {https://doi.org/10.1088/1361-648X/ac0c3c} {\bibfield  {journal} {\bibinfo
  {journal} {J. Phys.: Condens. Matter}\ }\textbf {\bibinfo {volume} {33}},\
  \bibinfo {pages} {405101} (\bibinfo {year} {2021})}\BibitemShut {NoStop}%
\bibitem [{\citenamefont {Krüger}\ and\ \citenamefont
  {Maes}(2016)}]{kruger_modified_2016}%
  \BibitemOpen
  \bibfield  {author} {\bibinfo {author} {\bibfnamefont {M.}~\bibnamefont
  {Krüger}}\ and\ \bibinfo {author} {\bibfnamefont {C.}~\bibnamefont {Maes}},\
  }\href {https://doi.org/10.1088/1361-648X/29/6/064004} {\bibfield  {journal}
  {\bibinfo  {journal} {J. Phys.: Condens. Matter}\ }\textbf {\bibinfo {volume}
  {29}},\ \bibinfo {pages} {064004} (\bibinfo {year} {2016})}\BibitemShut
  {NoStop}%
\bibitem [{\citenamefont {M\"uller}(2020)}]{muller_brownian_2020}%
  \BibitemOpen
  \bibfield  {author} {\bibinfo {author} {\bibfnamefont {B.}~\bibnamefont
  {M\"uller}},\ }\emph {\bibinfo {title} {Brownian Particles in Nonequilibrium
  Solvents}},\ \href
  {https://ediss.uni-goettingen.de/handle/21.11130/00-1735-0000-0005-12E6-3}
  {Ph.D. thesis},\ \bibinfo  {school} {{G}eorg-{A}ugust-{U}niversit{\"a}t
  {G}{\"o}ttingen} (\bibinfo {year} {2020})\BibitemShut {NoStop}%
\bibitem [{\citenamefont {Seifert}(2012)}]{seifert_stochastic_2012}%
  \BibitemOpen
  \bibfield  {author} {\bibinfo {author} {\bibfnamefont {U.}~\bibnamefont
  {Seifert}},\ }\href {https://doi.org/10.1088/0034-4885/75/12/126001}
  {\bibfield  {journal} {\bibinfo  {journal} {Rep. Prog. Phys.}\ }\textbf
  {\bibinfo {volume} {75}},\ \bibinfo {pages} {126001} (\bibinfo {year}
  {2012})}\BibitemShut {NoStop}%
\bibitem [{\citenamefont {Colangeli}\ \emph {et~al.}(2011)\citenamefont
  {Colangeli}, \citenamefont {Maes},\ and\ \citenamefont
  {Wynants}}]{colangeli_meaningful_2011}%
  \BibitemOpen
  \bibfield  {author} {\bibinfo {author} {\bibfnamefont {M.}~\bibnamefont
  {Colangeli}}, \bibinfo {author} {\bibfnamefont {C.}~\bibnamefont {Maes}},\
  and\ \bibinfo {author} {\bibfnamefont {B.}~\bibnamefont {Wynants}},\ }\href
  {https://doi.org/10.1088/1751-8113/44/9/095001} {\bibfield  {journal}
  {\bibinfo  {journal} {J. Phys. A: Math. Theor.}\ }\textbf {\bibinfo {volume}
  {44}},\ \bibinfo {pages} {095001} (\bibinfo {year} {2011})}\BibitemShut
  {NoStop}%
\bibitem [{\citenamefont {Basu}\ \emph {et~al.}(2015)\citenamefont {Basu},
  \citenamefont {Krüger}, \citenamefont {Lazarescu},\ and\ \citenamefont
  {Maes}}]{basu_frenetic_2015}%
  \BibitemOpen
  \bibfield  {author} {\bibinfo {author} {\bibfnamefont {U.}~\bibnamefont
  {Basu}}, \bibinfo {author} {\bibfnamefont {M.}~\bibnamefont {Krüger}},
  \bibinfo {author} {\bibfnamefont {A.}~\bibnamefont {Lazarescu}},\ and\
  \bibinfo {author} {\bibfnamefont {C.}~\bibnamefont {Maes}},\ }\href
  {https://doi.org/10.1039/C4CP04977B} {\bibfield  {journal} {\bibinfo
  {journal} {Phys. Chem. Chem. Phys.}\ }\textbf {\bibinfo {volume} {17}},\
  \bibinfo {pages} {6653} (\bibinfo {year} {2015})}\BibitemShut {NoStop}%
\bibitem [{\citenamefont {Maes}(2020)}]{maes_response_2020}%
  \BibitemOpen
  \bibfield  {author} {\bibinfo {author} {\bibfnamefont {C.}~\bibnamefont
  {Maes}},\ }\href
  {https://www.frontiersin.org/articles/10.3389/fphy.2020.00229} {\bibfield
  {journal} {\bibinfo  {journal} {Front. Phys.}\ }\textbf {\bibinfo {volume}
  {8}} (\bibinfo {year} {2020})}\BibitemShut {NoStop}%
\bibitem [{\citenamefont {Caspers}\ and\ \citenamefont
  {Krüger}(2024)}]{caspers_nonlinear_2024}%
  \BibitemOpen
  \bibfield  {author} {\bibinfo {author} {\bibfnamefont {J.}~\bibnamefont
  {Caspers}}\ and\ \bibinfo {author} {\bibfnamefont {M.}~\bibnamefont
  {Krüger}},\ }\href {https://doi.org/10.1063/5.0227674} {\bibfield  {journal}
  {\bibinfo  {journal} {J. Chem. Phys.}\ }\textbf {\bibinfo {volume} {161}},\
  \bibinfo {pages} {124109} (\bibinfo {year} {2024})}\BibitemShut {NoStop}%
\bibitem [{\citenamefont {Müller}\ \emph {et~al.}(2020)\citenamefont
  {Müller}, \citenamefont {Berner}, \citenamefont {Bechinger},\ and\
  \citenamefont {Krüger}}]{muller_properties_2020}%
  \BibitemOpen
  \bibfield  {author} {\bibinfo {author} {\bibfnamefont {B.}~\bibnamefont
  {Müller}}, \bibinfo {author} {\bibfnamefont {J.}~\bibnamefont {Berner}},
  \bibinfo {author} {\bibfnamefont {C.}~\bibnamefont {Bechinger}},\ and\
  \bibinfo {author} {\bibfnamefont {M.}~\bibnamefont {Krüger}},\ }\href
  {https://doi.org/10.1088/1367-2630/ab6a39} {\bibfield  {journal} {\bibinfo
  {journal} {New J. Phys.}\ }\textbf {\bibinfo {volume} {22}},\ \bibinfo
  {pages} {023014} (\bibinfo {year} {2020})}\BibitemShut {NoStop}%
\bibitem [{Note2()}]{Note2}%
  \BibitemOpen
  \bibinfo {note} {It will be detailed below why the expansion is performed
  around the equilibrium state corresponding to $x_t$.}\BibitemShut {Stop}%
\bibitem [{\citenamefont {Callen}\ and\ \citenamefont
  {Welton}(1951)}]{callen_irreversibility_1951}%
  \BibitemOpen
  \bibfield  {author} {\bibinfo {author} {\bibfnamefont {H.~B.}\ \bibnamefont
  {Callen}}\ and\ \bibinfo {author} {\bibfnamefont {T.~A.}\ \bibnamefont
  {Welton}},\ }\href {https://doi.org/10.1103/PhysRev.83.34} {\bibfield
  {journal} {\bibinfo  {journal} {Phys. Rev.}\ }\textbf {\bibinfo {volume}
  {83}},\ \bibinfo {pages} {34} (\bibinfo {year} {1951})}\BibitemShut {NoStop}%
\bibitem [{\citenamefont {Kubo}(1966)}]{kubo_fluctuation-dissipation_1966}%
  \BibitemOpen
  \bibfield  {author} {\bibinfo {author} {\bibfnamefont {R.}~\bibnamefont
  {Kubo}},\ }\href {https://doi.org/10.1088/0034-4885/29/1/306} {\bibfield
  {journal} {\bibinfo  {journal} {Rep. Prog. Phys.}\ }\textbf {\bibinfo
  {volume} {29}},\ \bibinfo {pages} {255} (\bibinfo {year} {1966})}\BibitemShut
  {NoStop}%
\bibitem [{\citenamefont {Maes}(2021)}]{maes_local_2021}%
  \BibitemOpen
  \bibfield  {author} {\bibinfo {author} {\bibfnamefont {C.}~\bibnamefont
  {Maes}},\ }\href {https://doi.org/10.21468/SciPostPhysLectNotes.32}
  {\bibfield  {journal} {\bibinfo  {journal} {SciPost Phys. Lect. Notes}\ ,\
  \bibinfo {pages} {32}} (\bibinfo {year} {2021})}\BibitemShut {NoStop}%
\bibitem [{\citenamefont {Baiesi}\ \emph {et~al.}(2009)\citenamefont {Baiesi},
  \citenamefont {Maes},\ and\ \citenamefont
  {Wynants}}]{baiesi_fluctuations_2009}%
  \BibitemOpen
  \bibfield  {author} {\bibinfo {author} {\bibfnamefont {M.}~\bibnamefont
  {Baiesi}}, \bibinfo {author} {\bibfnamefont {C.}~\bibnamefont {Maes}},\ and\
  \bibinfo {author} {\bibfnamefont {B.}~\bibnamefont {Wynants}},\ }\href
  {https://doi.org/10.1103/PhysRevLett.103.010602} {\bibfield  {journal}
  {\bibinfo  {journal} {Phys. Rev. Lett.}\ }\textbf {\bibinfo {volume} {103}},\
  \bibinfo {pages} {010602} (\bibinfo {year} {2009})}\BibitemShut {NoStop}%
\bibitem [{Note1()}]{Note1}%
  \BibitemOpen
  \bibinfo {note} {Local detailed balance amounts to assuming that all hidden
  degrees of freedom are equilibrated and in contact with heat baths. Consider,
  for example, a driven Brownian particle in a fluid. In a good approximation,
  the fluid molecules remain in thermal equilibrium~\cite {falasco_local_2021}.
  For explanations and counter-examples of systems where local detailed balance
  is broken, the reader is referred to Ref.~\cite
  {maes_local_2021}.}\BibitemShut {Stop}%
\bibitem [{\citenamefont {Holsten}\ and\ \citenamefont
  {Krüger}(2021)}]{holsten_thermodynamic_2021}%
  \BibitemOpen
  \bibfield  {author} {\bibinfo {author} {\bibfnamefont {T.}~\bibnamefont
  {Holsten}}\ and\ \bibinfo {author} {\bibfnamefont {M.}~\bibnamefont
  {Krüger}},\ }\href {https://doi.org/10.1103/PhysRevE.103.032116} {\bibfield
  {journal} {\bibinfo  {journal} {Phys. Rev. E}\ }\textbf {\bibinfo {volume}
  {103}},\ \bibinfo {pages} {032116} (\bibinfo {year} {2021})}\BibitemShut
  {NoStop}%
\bibitem [{\citenamefont {Onsager}\ and\ \citenamefont
  {Machlup}(1953)}]{onsager_fluctuations_1953}%
  \BibitemOpen
  \bibfield  {author} {\bibinfo {author} {\bibfnamefont {L.}~\bibnamefont
  {Onsager}}\ and\ \bibinfo {author} {\bibfnamefont {S.}~\bibnamefont
  {Machlup}},\ }\href {https://doi.org/10.1103/PhysRev.91.1505} {\bibfield
  {journal} {\bibinfo  {journal} {Phys. Rev.}\ }\textbf {\bibinfo {volume}
  {91}},\ \bibinfo {pages} {1505} (\bibinfo {year} {1953})}\BibitemShut
  {NoStop}%
\bibitem [{\citenamefont {Falasco}\ and\ \citenamefont
  {Esposito}(2021)}]{falasco_local_2021}%
  \BibitemOpen
  \bibfield  {author} {\bibinfo {author} {\bibfnamefont {G.}~\bibnamefont
  {Falasco}}\ and\ \bibinfo {author} {\bibfnamefont {M.}~\bibnamefont
  {Esposito}},\ }\href {https://doi.org/10.1103/PhysRevE.103.042114} {\bibfield
   {journal} {\bibinfo  {journal} {Phys. Rev. E}\ }\textbf {\bibinfo {volume}
  {103}},\ \bibinfo {pages} {042114} (\bibinfo {year} {2021})}\BibitemShut
  {NoStop}%
\end{thebibliography}

\end{document}